\documentclass[final,1p,times]{elsarticle}

\usepackage[table]{xcolor}
\usepackage{amssymb}
\usepackage{xfrac}
\usepackage{amsmath}
\usepackage{todonotes}
\usepackage{tabularx}
\usepackage{booktabs}
\usepackage{multirow}
\usepackage{url}
\usepackage{hyperref}
\usepackage{rotating}
\usepackage{color}
\usepackage{colortbl}
\usepackage{wrapfig}
\usepackage{subcaption}
\usepackage{longtable}
\usepackage{todonotes}

\definecolor{greenBack}{rgb}{ .796,  .875,  .812}
\definecolor{greenText}{rgb}{ 0,  .38,  0}
\definecolor{orangeBack}{rgb}{ .973,  .929,  .831}
\definecolor{orangeText}{rgb}{ .612,  .341,  0}
\definecolor{redBack}{rgb}{ .937,  .78,  .78}
\definecolor{redText}{rgb}{ .612,  0,  .024}
\definecolor{dynBlue}{rgb}{ .106,  .455,  .569}

\immediate\write18{%
makeindex -s nomencl.ist -o \jobname.nls -t \jobname.nlg \jobname.nlo%
}
\usepackage{framed} 
\usepackage{multicol} 
\usepackage{nomencl} 
\makenomenclature
\renewcommand*\nompreamble{\begin{multicols}{2}}
\renewcommand*\nompostamble{\end{multicols}}

 \usepackage{lineno}

\journal{Mechanical Systems and Signal Processing}

\begin{document}

\begin{frontmatter}



\title{Deep learning for brake squeal: brake noise detection, characterization and prediction}


\author[dyn]{Merten Stender\corref{cor1}}
\author[audi]{Merten Tiedemann}
\author[hm]{David Spieler}
\author[dyn,ifpt]{Daniel Schoepflin}
\author[dyn,ic]{Norbert Hoffmann}
\author[dyn,uts]{Sebastian Oberst}

\address[dyn]{Dynamics Group, Hamburg University of Technology, Hamburg, Germany}
\address[audi]{Brake Systems Development, AUDI AG, Ingolstadt, Germany}
\address[hm]{Department of Computer Science and Mathematics, University of Applied Sciences Munich, Germany}
\address[ifpt]{Institute for Aircraft Production Technology, Hamburg University of Technology, Hamburg, Germany}
\address[ic]{Department of Mechanical Engineering, Imperial College London, London, United Kingdom}
\address[uts]{Centre for Audio, Acoustics and Vibration, Faculty of Engineering and Information Technology, University of Technology Sydney, Sydney, Australia}

\cortext[cor1]{Corresponding author}

\begin{abstract}

Despite significant advances in modeling of friction-induced vibrations and brake squeal, the majority of industrial research and design is still conducted experimentally, since many aspects of squeal and its mechanisms involved remain unknown. In practice, measurement data is available in large amounts. We report here for the first time on novel strategies for handling data-intensive vibration testings to gain better insights into friction brake system vibrations and noise generation mechanisms. Machine learning-based methods to detect and characterize vibrations, to understand sensitivities and to predict brake squeal are applied with the aim to illustrate how interdisciplinary approaches can leverage the potential of data science techniques for classical mechanical engineering challenges.

In the first part, a deep learning brake squeal detector is developed to identify several classes of typical friction noise recordings. The detection method is rooted in recent computer vision techniques for object detection based on convolutional neural networks (CNN). It allows to overcome limitations of classical approaches that solely rely on instantaneous spectral properties of the recorded noise. Results indicate superior detection and characterization quality when compared to a state-of-the-art brake squeal detector. In the second part, a recurrent neural network (RNN) is employed to learn the parametric patterns that determine the dynamic stability of an operating brake system. Given a set of multivariate loading conditions, the RNN learns to predict the noise generation of the structure. The validated RNN represents a virtual twin model for the squeal behavior of a specific brake system. It is found that this model can predict the occurrence and the onset of brake squeal with high accuracy and that it can identify the complicated patterns and temporal dependencies in the loading conditions that drive the dynamical structure into regimes of instability. Large data sets from commercial brake system testing are used to train and validate the models. \emph{This work is a contribution to the MSSP Special Issue in Honor of Professor Lothar Gaul.}

\end{abstract}


\end{frontmatter}

\begin{table*}[!t]   
\begin{framed}
\nomenclature{AP}{average precision}
\nomenclature{AUC}{area under curve}
\nomenclature{$C$}{object detection confidence score}
\nomenclature{CNN}{convolutional neural network}
\nomenclature{CV}{computer vision}
\nomenclature{$\mathfrak{C}$}{classifier}
\nomenclature{DL}{deep learning}
\nomenclature{$d_{\mathrm{sq}}$}{brake squeal sound duration}
\nomenclature{FFT}{fast Fourier transform}
\nomenclature{FP}{false positives}
\nomenclature{FN}{false negatives}
\nomenclature{FIV}{friction-induced vibrations}
\nomenclature{$f_{\mathrm{sq}}$}{brake squeal sound frequency}
\nomenclature{$f_{\mathrm{s}}$}{sampling rate}
\nomenclature{$F_1$}{$F_1$ classification metric}
\nomenclature{$H$}{relative humidity}
\nomenclature{$\mathbf{h}$}{memory state}
\nomenclature{$h$}{shift parameter (sliding window)}
\nomenclature{IoU}{intersection over union}
\nomenclature{$m$}{number of sensors}
\nomenclature{$\lambda$}{complex eigenvalue}
\nomenclature{LSTM}{long-short-term memory (network)}
\nomenclature{$l_{\mathrm{sq}}$}{squeal sound pressure level}
\nomenclature{$\tilde{m}$}{number of actual system parameters}
\nomenclature{$M$}{braking torque}
\nomenclature{MCC}{Matthews correlation coefficient}
\nomenclature{mAP}{mean average precision}
\nomenclature{$\mu$}{friction coefficient}
\nomenclature{$n_{\mathrm{t}}$}{number of time series samples}
\nomenclature{$N$}{number of brakings}
\nomenclature{$N_{\mathrm{sq}}$}{number of squealing brakings}
\nomenclature{$N^{\mathrm{win}}$}{number of samples after windowing}
\nomenclature{$N^{\mathrm{win}}_{\mathrm{sq}}$}{number of squealing samples after windowing}
\nomenclature{NVH}{noise, vibration, harshness}
\nomenclature{$\Omega$}{rotational velocity}
\nomenclature{$p$}{brake line pressure}
\nomenclature{PRC}{precision recall curve}
\nomenclature{ReLu}{rectified linear unit}
\nomenclature{RNN}{recurrent neural network}
\nomenclature{$\mathbf{s}$}{univariate time series}
\nomenclature{$\mathbf{S}$}{multivariate time series}
\nomenclature{SP(L)}{sound pressure (level)}
\nomenclature{$\sigma$}{activation function}
\nomenclature{$t$}{time}
\nomenclature{$T_{\mathrm{rot}}$}{disk temperature}
\nomenclature{$T_{\mathrm{fluid}}$}{braking fluid temperature}
\nomenclature{$T_{\mathrm{amb}}$}{ambient temperature}
\nomenclature{TP}{true positives}
\nomenclature{TN}{true negatives}
\nomenclature{$W$}{network weights}
\nomenclature{$\xi$}{(bifurcation) parameter}
\nomenclature{$x$, $z$}{bounding box locations}
\nomenclature{$\left( \, \tilde{\cdot} \,\right)$}{prediction}
\nomenclature{$y$}{class label to predict}
\nomenclature{$Y$}{multivariate output sequence}
\printnomenclature
\end{framed}
\end{table*}

\section{Introduction}

Noise, vibration and harshness (NVH) issues in friction brakes are one of the most relevant customer claims in the automotive industry and also omnipresent in rail vehicles \cite{Kinkaid.2003, Brunel.2006,Hoffmann.2008,Sinou.2013}. During the last two decades, research activities, cf. Figure~\ref{fig:papers}~(a), have addressed fundamental instability mechanisms \cite{Hoffmann.2002,Kruse.2015b,Wagner.2007}, design countermeasures \cite{Lazzari.2018,Stender.2016b}, advanced computational modeling \cite{Ouyang.2005,Massi.2007,Coudeyras.2009c,Brunetti.2016, Stender.2018}, signal analysis \cite{Oberst.2011,Wernitz.2012,Vitanov.2014,Oberst.2015,Stender.2019b,Stender.2019c}, uncertainty analysis \cite{Oberst.2011b,Hanselowski.2014,Zhang.2016c,Renault.2016} and many more using experimental data and numerical models. However, until today the model prediction quality is mostly unsatisfactory \cite{Butlin.2010, Nobari.2015b}. The understanding of the actual instantaneous system conditions, responsible for self-excitation mechanisms of real-world braking applications, is only very limited. Decades of brake squeal research illustrate the fugitive character of the highly nonlinear, multi-scale and potentially chaotic friction-induced vibrations (FIV) \cite{Oberst.2011, Butlin.2010, Fu.2015, Stender.2019f}. 

Whenever systems are studied experimentally \cite{Giannini.2006,Akay.2009, Sinou.2019}, the occurrence of NVH-related vibrations and noise generation mechanisms has to be monitored. While the detection of large-amplitude oscillations in sensor measurements of a brake system may seem trivial, the automatic detection and classification of typical brake sounds is still challenging. Identification of vibrations and noise in measurements plays a crucial role in both academia and industry: in the era of big data and deep learning \cite{LeCun.2015b} large amounts of data are recorded which rule out manual approaches. Cost-intensive design decisions, tailored research approaches and countermeasures are based on the NVH assessment of a brake system, and thus on the NVH events encountered during testing. Higher detection quality and accuracy is therefore of fundamental interest. To reduce FIV in brake systems, unstable regimes must be identified. From an energetic perspective, the friction interface provides energy input to the structure composed of multiple parts and assembled by mechanical joints \cite{Gaul.2001,Hetzler.2014,Brake.2018} which add significant amounts of damping and nonlinearity \cite{Tiedemann.2015b,Grabner.2014}. Furthermore, the structure experiences external loads, such as excitation from the road, and changing environmental conditions, such as changing ambient temperatures. Stability, and thus the vibration behavior of the structure, is governed by the flow, localization and balance of energy between sources and sinks. Due to temperature-dependent material parameters, non-linear force-displacement characteristics and ever-changing friction interfaces conditions \cite{Hetzler.2012, Cabboi.2018}, local and global stiffness and damping parameters vary between and during brake operations, leading to changing system stability \cite{Butlin.2009}. As a result, the structure may exhibit rich and intricate bifurcation behavior with respect to one or multiple parameters \cite{Kruse.2015b, Jahn.2019b, Stender.2020b}. Given the complexity of the system, one cannot assume a single parameter, such as the relative sliding velocity of the brake pad, to generally initiate instability and to dominate the overall bifurcation behavior. Instead, combinations of various parameters, their instantaneous changes as well as their historical values, do drive the system dynamics. The rich corpus of research on frictional systems has identified various mechanisms for instability \cite{Tonazzi.2013}, such as stick-slip \cite{Tonazzi.2013, Liu.2019b}, a negative friction slope \cite{Lacerra.2018,Ouyang.1998, Hetzler.2007} and mode-coupling \cite{Hoffmann.2002, Wagner.2007,Hoffmann.2003}. However, most of these mechanisms represent idealized  single-point contact systems. Spatially distributed contacts are well-known for adding aspects of synchronization \cite{Marszal.2016}, multi-stability \cite{Stender.2020b, Ryabov.1995, Papangelo.2017} and localization \cite{Papangelo.2018, Papangelo.2019}, i.e. complicating the picture of stability drastically. Previous work \cite{Stender.2019b} on extracting the instantaneous growth rates from experimental data, i.e. quantifying linear stability and instantaneous damping, illustrates the sensitivity of the effective damping with respect to loading conditions. As a result, and along with the aforementioned parametric changes to the system properties, it remains a difficult task to identify and understand the underlying parameter space and conditions that drive a real brake system into instability.

\subsection{System understanding}

\begin{figure}[htb]
\centering
\includegraphics[width=0.98\columnwidth]{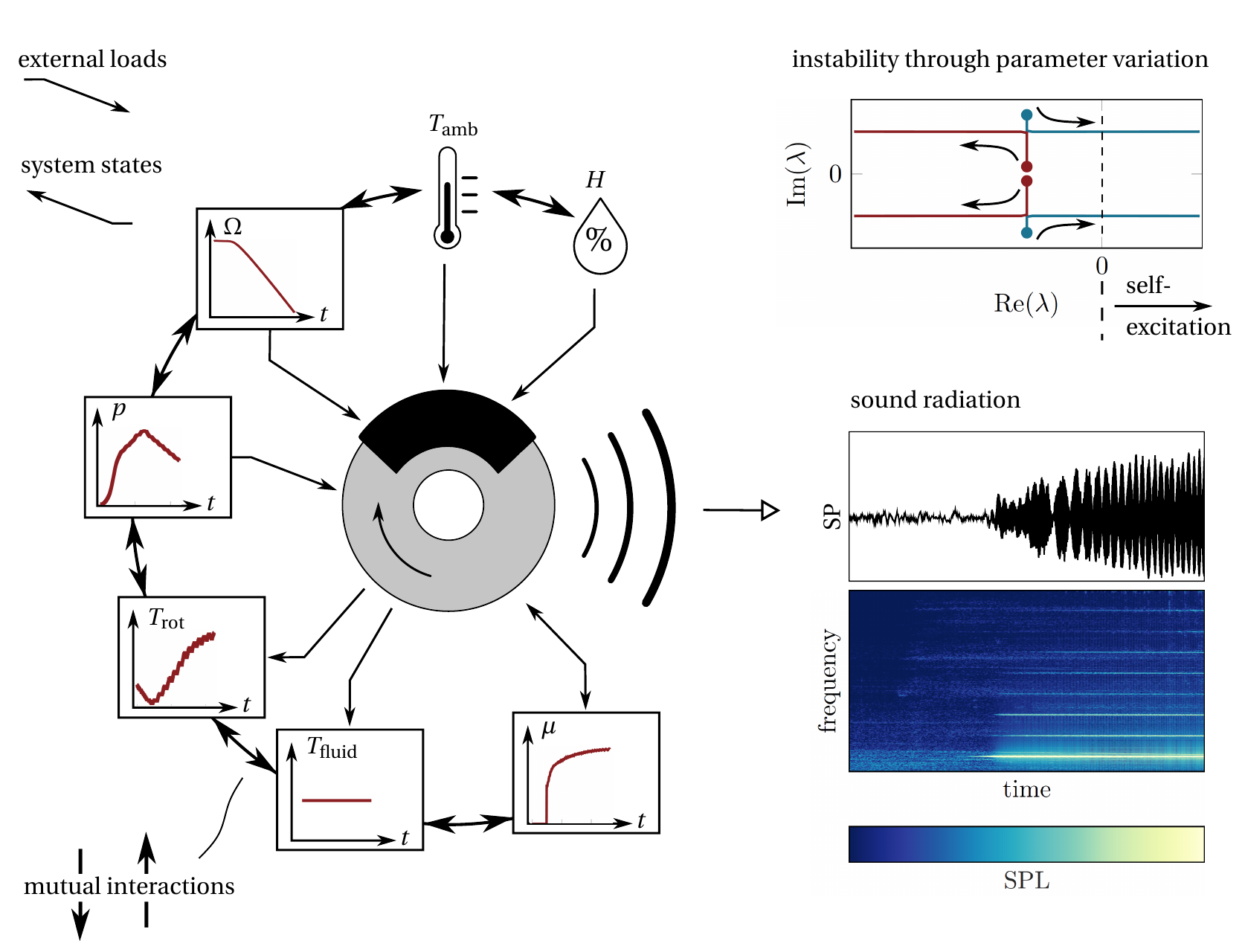}%
\caption{An example of a single brake stop and  measurements of  disk rotation $\Omega$,  brake line pressure $p$,  ambient temperature $T_{\mathrm{amb}}$ and  relative air humidity $H$. As a consequence, other system parameters change, such as the disk surface temperature $T_{\mathrm{rot}}$, the brake fluid temperature $T_{\mathrm{fluid}}$ and the friction coefficient $\mu$. These parameter variations can render the system unstable and let friction-induced vibrations grow until eventually audible noise, especially squeal, is emitted.}
\label{fig:system_understanding}%
\end{figure}

Figure~\ref{fig:system_understanding} visualizes various aspects of multi-physic loads, parameter variations, self-excitation and transient noise encountered during brake operation. The mechanical components interact with the driver's deceleration request and internal states like the friction coefficient and component temperatures change. These parametric changes can cause instability of the steady sliding state, leading to non-stationary, high-intensity and high-frequency non-equilibrium dynamics. From a classical modeling perspective that is concerned with stability analysis by complex eigenvalues, the parameter variation causes positive real parts that destabilize the equilibrium state. A detailed time-frequency analysis of the vibration measurement allows to characterize the system response qualitatively and quantitatively. Figure~\ref{fig:system_understanding} displays the actual loads recorded on a dynamometer, the emitted sound pressure (SP) recorded by a microphone, and the Gabor transform of the A-weighted sound pressure level (SPL) for a single brake operation. It becomes obvious that slow processes, i.e. the loads, as well as fast processes, i.e. interface dynamics, vibrations and sound emissions, are involved in the brake dynamics. As the loads interact with each other, and with the structure itself, we consider the overall dynamical system to complex owing to its multi-physic, multi-scale, and non-stationary characteristics. 

\subsection{Data from brake system testing}

\begin{figure}[ht]
\centering
\includegraphics[width=0.999\columnwidth]{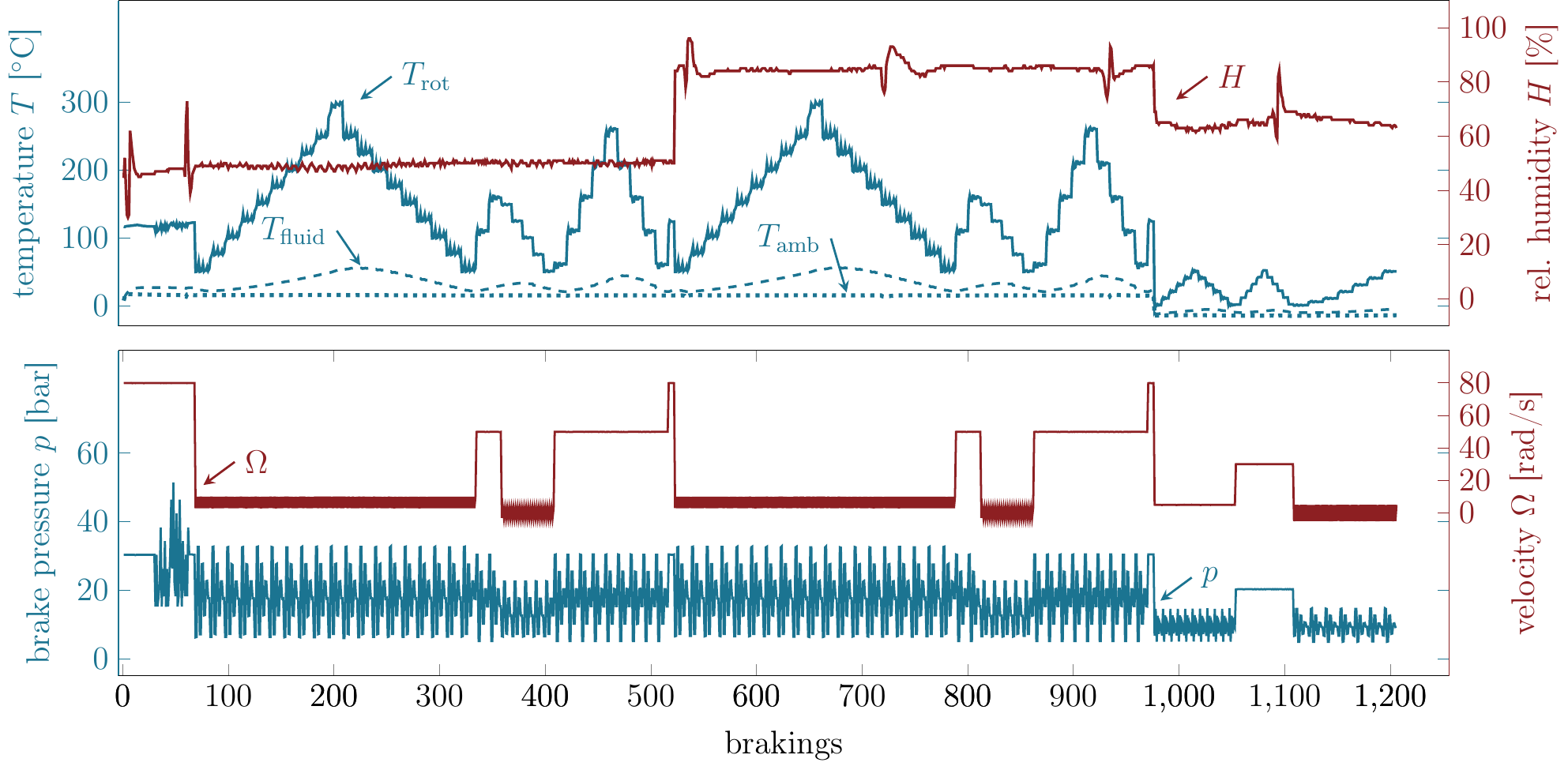}%
\caption{A complete brake system NVH test derived from the SAE-J2521 protocol. The test matrix involves several temperature ramps, brakings at various pressure and velocity levels and varying air humidity values. The maximum values of each measurement channel are reported per braking}
\label{fig:data_set_viz}%
\end{figure}

The data sets used in this work have been recorded according to test protocols derived from the SAE-J2521 procedure \cite{SAEJ.2521}. Quarter-car sections are subjected to the protocol on a dynamometer test bench. A matrix test is conducted for variations of the type of braking (stop or drag braking), different  durations and various combinations of operational parameters, such as rotational velocity, brake line pressure or ambient air temperature. For running-in and conditioning the system, 30 brake stop are conducted in the beginning of each test run. Then, nine temperature ramps at several pressure and velocity levels are performed. Overall, the protocol is designed to cover a wide range of brake scenarios. Figure~\ref{fig:data_set_viz} depicts the evolution of three temperature measurements (rotor $T_{\mathrm{rot}}$, ambient $T_{\mathrm{amb}}$ and fluid $T_{\mathrm{fluid}}$), the initial rotor velocity ($\Omega$) and the brake pressure ($p$) values for a complete test run of 1206 brakings. The time evolution of the loading parameters is recorded in the form of time series data sampled at $f_{\mathrm{s}}=100$\,Hz. A microphone located in proximity to the brake disc with a sampling frequency of $f_{\mathrm{s}}=51.2$\,kHz monitors the development of brake noise. Typical sequences measured during a single stop braking are displayed in Figure~\ref{fig:system_understanding}.

\subsection{Objectives and structure of this work}
Given the abundance of data in the domain of friction-induced vibrations and brake noise, we were interested in better understanding the potentials and limitations of applying recent data-driven approaches that are nowadays omnipresent in many fields science and technology \cite{LeCun.2015}. However, so far there has been no attempt to leverage the potential of machine learning to study friction-induced noise of brake systems - which we will do within this study for the first time by illustrating pathways for data-driven treatment of friction-induced vibrations which will hopefully be fruitful in fostering further research in this field.

The first part of this study is concerned with a novel brake noise detector for NVH applications in brake systems to locate and characterize friction phenomena using microphone recordings. The design of a data-driven virtual twin for the brake behavior is presented in the second part, where the loading sequences are used as input to predict the vibrational response of brake systems. The noise detector from the first part finds here its application to label large data sets in a completely automated fashion.

\section{Part 1: Vibration detection and characterization}

In NVH engineering, friction noise phenomena such as \textit{judder}, \textit{creep-groan} and \textit{squeal}, which are either forced, self-excited and combinations thereof, are well-known \cite{Akay.2002}. Brake squeal is considered one of most pressing issue in NVH departments of the automotive brake industry \cite{Akay.2002, Abendroth.2000, AbuBakar.2006}. However, signals recorded during experimental testing are multi-scale and transient \cite{Oberst.2011, Stender.2019f}. The difficulties to detect FIV-triggered noise in sound recordings arise from distortion and electronic noise in the signal, and environmental noise contamination originating for instance from the dynamometer, the motor, the ventilation system and other external engines, which steadily contribute with a colored noise floor and aggregate-dependent periodicities. The operation of the brake system itself creates additional dynamics due to the rotation of an imperfectly flat disk, piston actuation and resulting large scale motion of the complete system. During vehicle operation in the field, the contribution of other sources to the overall signal level becomes stronger e.g. due to the excitation of the street (high dimensional, quasi-random) or potential interactions (vibration, acoustics) from other traffic participants.

At the same time, a wide range of scientific disciplines shares the very general objective of finding characteristic patterns in audio or vibration recordings. Some of these disciplines have already adopted deep learning techniques: In \cite{Lavner.2016} baby cries were identified in noise-contaminated recordings using  deep convolutional neural networks (CNN) which outperformed  a classical handcrafted detector. CNNs have been employed to classify bird calls making used of more than $35.000$ recordings of $1500$ bird species; raw audio signals in combination with their visual representations in the form of spectrograms showed to produce high detection rates \cite{Kahl.2017}. Other examples include the acoustic detection of flying drones \cite{Jeon.1202017}, audio signal analysis for sleep quality assessment \cite{Amiriparian.2017} or the detection of scream and gunshots
\cite{Valenzise.2007,Laffitte.2016}. 

In structural dynamics related to FIV, highly-automated procedures for detection and classification are often limited to amplitude-based criteria and spectral methods using fast Fourier transforms \cite{Mauer.2007} which work well for tonal noise, however, which may fail in case of a more complex signal character \cite{Oberst.2018f, Oberst.2017b}. However, experienced engineers are often very well able to visually identify different noises in a case-by-case manner using spectrograms derived from the sound measurements. Hence, we propose a highly automatic technique to detect and classify different brake noises by applying deep learning techniques which have been used for object detection in images \cite{LeCun.2015}. 

This part is organized as follows: first, state-of-the-art squeal detection approaches, their limitations and deep learning computer vision techniques are re-visited. Next, a novel noise detection algorithm based on deep learning (DL) is proposed which is applied to a large set of automotive disk brake sound measurements. Noise events are classified, and in a second step localised using noise classes in the frequency and in the time domain.  The performance of this novel technique is compared to classical approaches. This study uses microphone measurements as input signals. Naturally, also vibration measurements from acceleration sensors can be used with very little changes (only related to different value ranges of the physical quantities) to the algorithms.

\subsection{Requirements for NVH detectors}
\label{sec:requirements_detector}

\begin{figure}[ht]
\centering
\includegraphics[width=0.98\columnwidth]{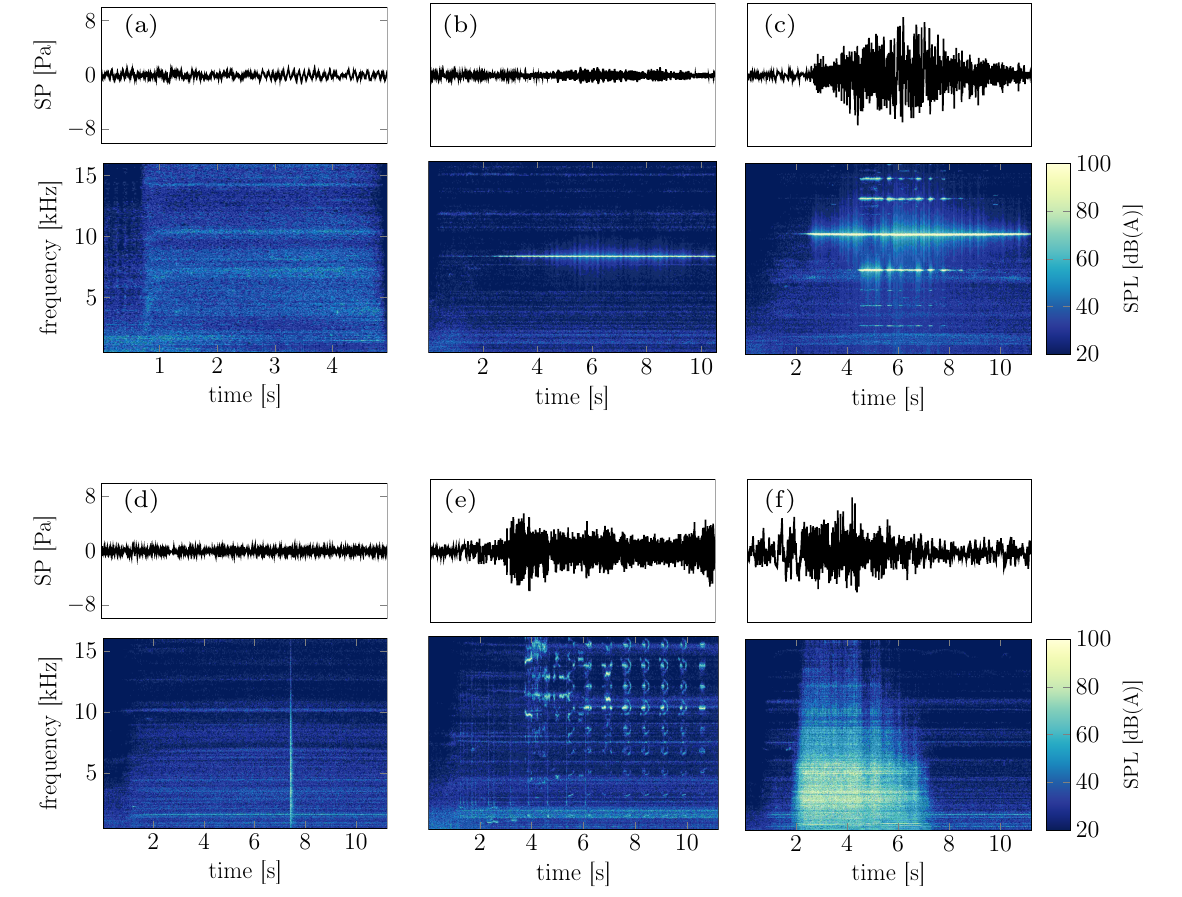}%
\caption{Microphone recording of typical brake noise recorded on a dynamometer: (a) quiet braking, (b) monofrequent  squeal, (c) multiple co-existing squeal events, (d) click sound of the pad, (e)  wirebrush noise involving multiple short impulses and chirps, and (f) broad-band noise artefacts}
\label{fig:phenomena}%
\end{figure}

To detect brake noise successfully, the dominant characteristics need to be extracted, which are, however, most often buried within noisy, transient or multi-scale measurements  \cite{Parthasarathy.2006}. We are aiming at developing a noise detector that, given a single microphone recording, can 
\begin{enumerate}
    \item classify brake noise recordings into categories (\textit{'Is there a brake sound in this recording?'}), and
    \item characterize different classes of noise in terms of their frequency content and duration \textit{'At which frequency and when does the noise occur?'} with a pre-defined frequency and time resolution. 
\end{enumerate}

In this work, four prototypical types of noise are considered which can be correlated to different vibration states. Particularly, these noise events are
\begin{itemize}
\item \textbf{squeal}: tonal sound in the frequency range $1 \leq f_{\mathrm{sq}} \leq 16$\,kHz with amplitudes above $l_{\mathrm{sq}}=50$\,dB(A). Multiple squeals at different frequency ranges and/or time instances, including higher harmonics, are possible.
\item \textbf{click sound} of the pad with an impulse-like characteristic that spans the complete frequency range,
\item \textbf{wirebrush} involving many short-time sounds at different frequencies, and
\item \textbf{noise artefacts} with a broad-banded frequency signature and high amplitudes.
\end{itemize}

 Examples of different signal types covering the four classes are depicted in Figure~\ref{fig:phenomena} to illustrate their spectral characteristics, but also to illustrate the plethora and complexity of friction-induced noise of brake systems.

\subsection{Conventional spectral squeal detection}

Only few references can be identified that explain current approaches to detect the class of brake squeal sounds in recordings from microphones or acceleration sensors \cite{Mauer.2007,VDA305,BONI.2006}. Conceptually, the tonal character of squeal sounds restricts the vibration energy to being confined to a narrow frequency range, i.e. a sharp, dominant and weakly damped peak in the signal’s Fourier transform. Typically, several conditions are posed on the amplitude, sharpness and duration of the vibration event to be identified as a squeal sound \cite{VDA305}.  
\begin{figure}[ht]
\centering
\includegraphics[width=0.9\columnwidth]{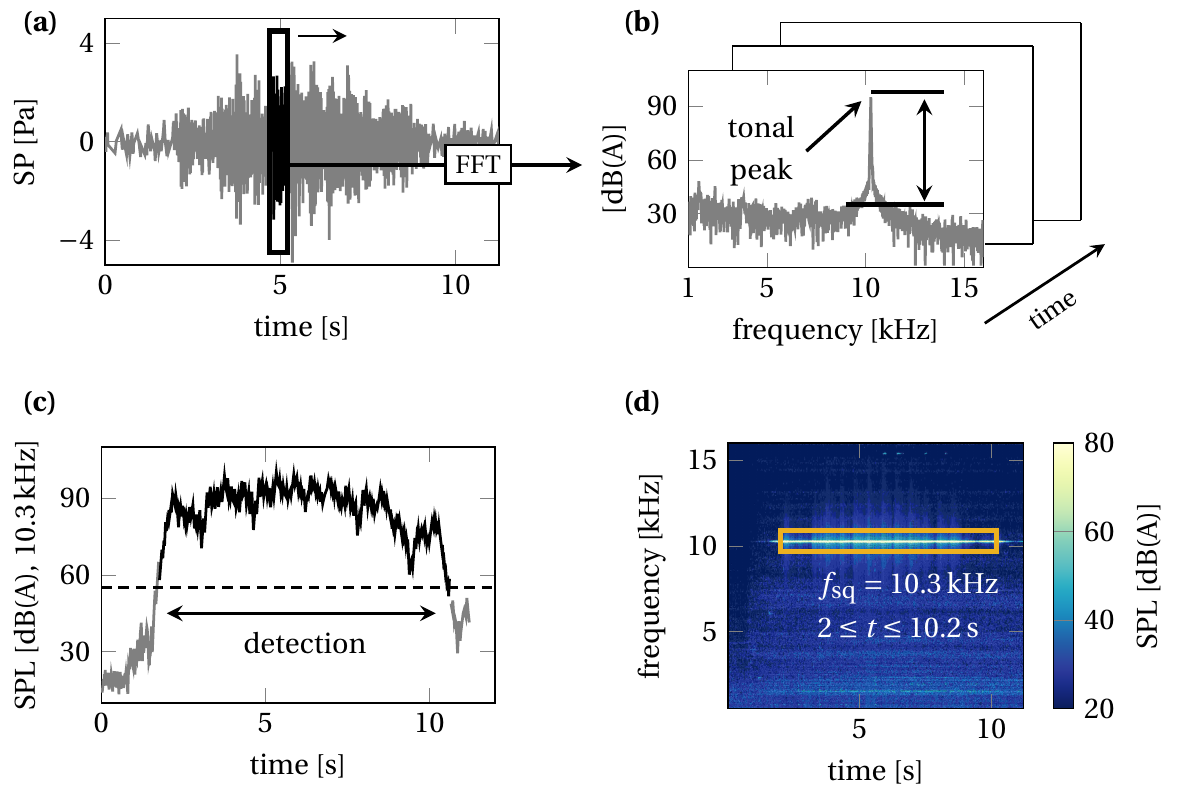}%
\caption{Schematic of a  squeal noise detector based on commonly used Fourier spectra: (a) microphone measurement of the sound pressure (SP) as recorded during braking. Using a sliding window approach, a series of spectra is computed to search for tonal events. (b) a tonal event is defined by the sharpness of the peak, i.e. its height above the surrounding frequency bandwidth. (c) the sound pressure level of candidate squeal frequencies $f_{\mathrm{sq}}$ is tracked over time to find the squeal duration and the squeal amplitude satisfying a minimal level such as $50$\,dB(A). The final squeal detection result is depicted in the spectrogram (d)}%
\label{fig:squealDetection}%
\end{figure}
For comparison to our neural network based classifier, we implemented such a spectral squeal noise detector for this work. The spectral detector is schematised in Figure~\ref{fig:squealDetection}. Using a sliding window, the amplitude spectra of successive epochs of the signal are computed, and peaks are detected to assess their sharpness. The tonal character of a peak is confirmed if its amplitude exceeds a threshold value above the mean level of a $1$\,kHz frequency band centered at the candidate's peak frequency. As squeal frequencies may slightly shift over time, peaks are grouped to a centre frequency if they do not differ by more than $\pm 100$\,Hz. For each centre frequency, the duration of a tonal sound event above the critical sound pressure level, for example $50$\,dB(A), is evaluated. Single short events are discarded while interrupted sounds of the same frequency are combined to a single event. 

In the example displayed in Figure~\ref{fig:squealDetection}, the dominant squeal event can be detected easily based on the aforementioned spectral properties. However, the presence of simultaneously occurring different friction-induced vibration phenomena render the spectral squeal detection impractical, and limit the spectral detector to only squeal noise. Also, multiple co-existing tonal sounds, frequency and amplitude modulations, superposed broad-band noise and contamination from the testing environment can hinder a robust peak detection resulting in erroneous classification in form of false positives (FP: squeal detected even though there is none) and false negatives (FN: no squeal detected even though there is one) counts\footnote{false negatives commonly represent an error of the 2nd kind which is generally to be avoided by all means.}.

\subsection{Object detection in computer vision}

\begin{figure}[ht]
\includegraphics[width=1.0\columnwidth]{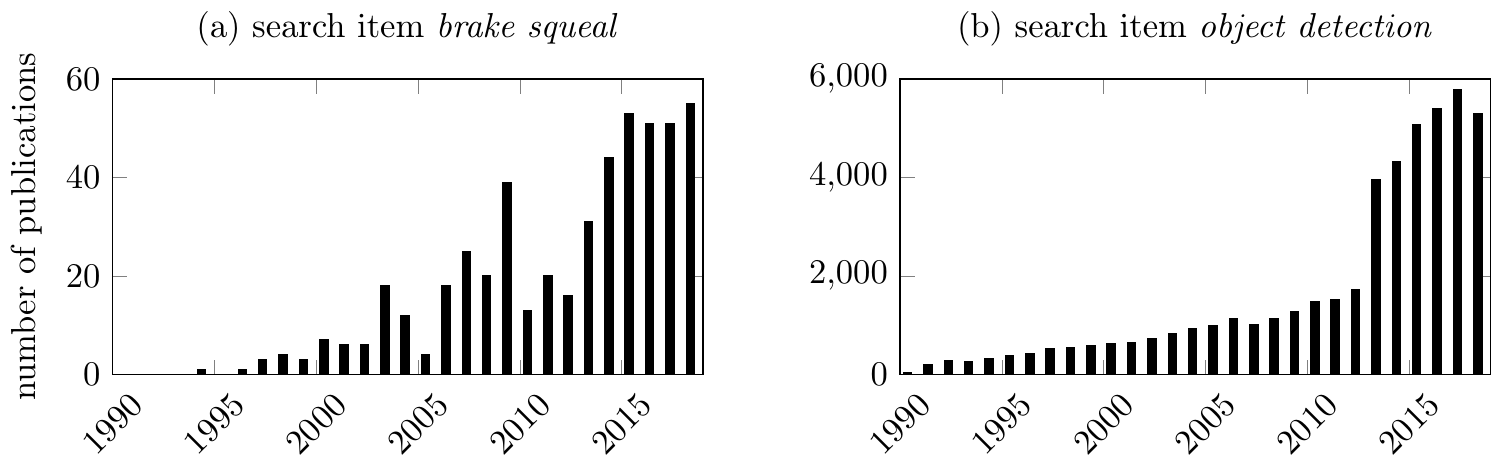}%
\caption{Number of new publications per year for search items (a) \textit{brake squeal} and (b) \textit{object detection} from Web of Science. Search results include articles, proceedings, reviews and book chapters, as per May 13th 2019}
\label{fig:papers}%
\end{figure}

Brake noise and object detection have been traditionally studied by different research communities as indicted by Figure~\ref{fig:papers}. While the increasing number of works on brake squeal indicate the growing demand for comfort in the automotive industry, deep learning (DL) has been a disruptive element for object detection interfacing many more areas of research with a major jump in publication numbers\footnote{Despite the global trend of increasing numbers of publications, we think that this point truly reflects the impact of DL on computer vision} in DL object detection beyond 2012. Before we illustrate our novel approach to combine those two disciplines, we shortly re-visit the core methodologies, opportunities and limitations of state-of-the-art DL object detection procedures. 

Computer vision tasks can be categorized into image classification, i.e. labeling an image, and object detection, i.e. labeling one or multiple objects within an image. Prior to a broad application of artificial neural networks (see Figure~\ref{fig:papers}~(b) before 2012), classical computer vision (CV) improved steadily by the use of increasingly complex handcrafted features. While \cite{Viola.2001} used cascades of low-level features for face detection, \cite{Dalal.2005} introduced more complex oriented gradients for shape detection. Next, \cite{Felzenszwalb.2010} proposed deformable templates and \cite{Dollar.2014} used multi-resolution image features for better object detection. The annual ImageNet \cite{Russakovsky.2015} and PASCAL \cite{Everingham.2015} challenge for image classification and object localization illustrated the incremental improvements of conventional CV techniques using increasingly complex features and classifiers. In 2012, first neural network-based approaches (AlexNet \cite{Krizhevsky.2012}) showcased their performance and flexibility in performing object detection tasks. Since then, deep learning methods dominate in those competitions owing to a substantially higher performance \cite{Zhao.2019b}. 

In this work, the object detection methods are used. Given an input image, the detector needs to frame detected objects with so-called bounding boxes and assign a label to each box. Two major approaches exist for building such detectors: region proposal-based and regression-based methods \cite{Zhao.2019b}. The first branch is a two-step process including the task of bounding box regression (i.e. optimizing the correct size and position of the object's bounding box) and the task of object classification. Region proposals are generated first, and the classification of each proposal into a category is performed second. Region (R) proposal based methods with convolutional neural networks (CNN) developed along R-CNN \cite{Girshick.2014}, Fast R-CNN \cite{Girshick.2015}, Faster R-CNN \cite{Ren.2015}, R-FCN (region-based fully connected networks) \cite{Dai.2016} and others. The second branch of methods achieves the regression task and the classification task at once, therefore greatly reducing the computational efforts. All of the aforementioned models rely on convolutional layers as core building blocks for the network. However, several different network configurations are well-established nowadays. The \textit{AlexNet} \cite{Krizhevsky.2012} is generally composed of five convolutional and three fully connected (FC) layers using rectified linear units (ReLu) as activation function. To reduce complexity, the \textit{inception} module of GoogLeNet \cite{Szegedy.2015} relies on sparse convolutional layers and simpler pooling layers instead of FC layers. Residual networks (\textit{ResNet}) avoid the vanishing gradient problem and degradation by a modular structure that reduces the network error faster. We will use instances of the inception and ResNet models in the following studies.

\subsection{Deep learning brake noise detection}
	
\begin{figure}[htb]
\includegraphics[width=0.97\columnwidth]{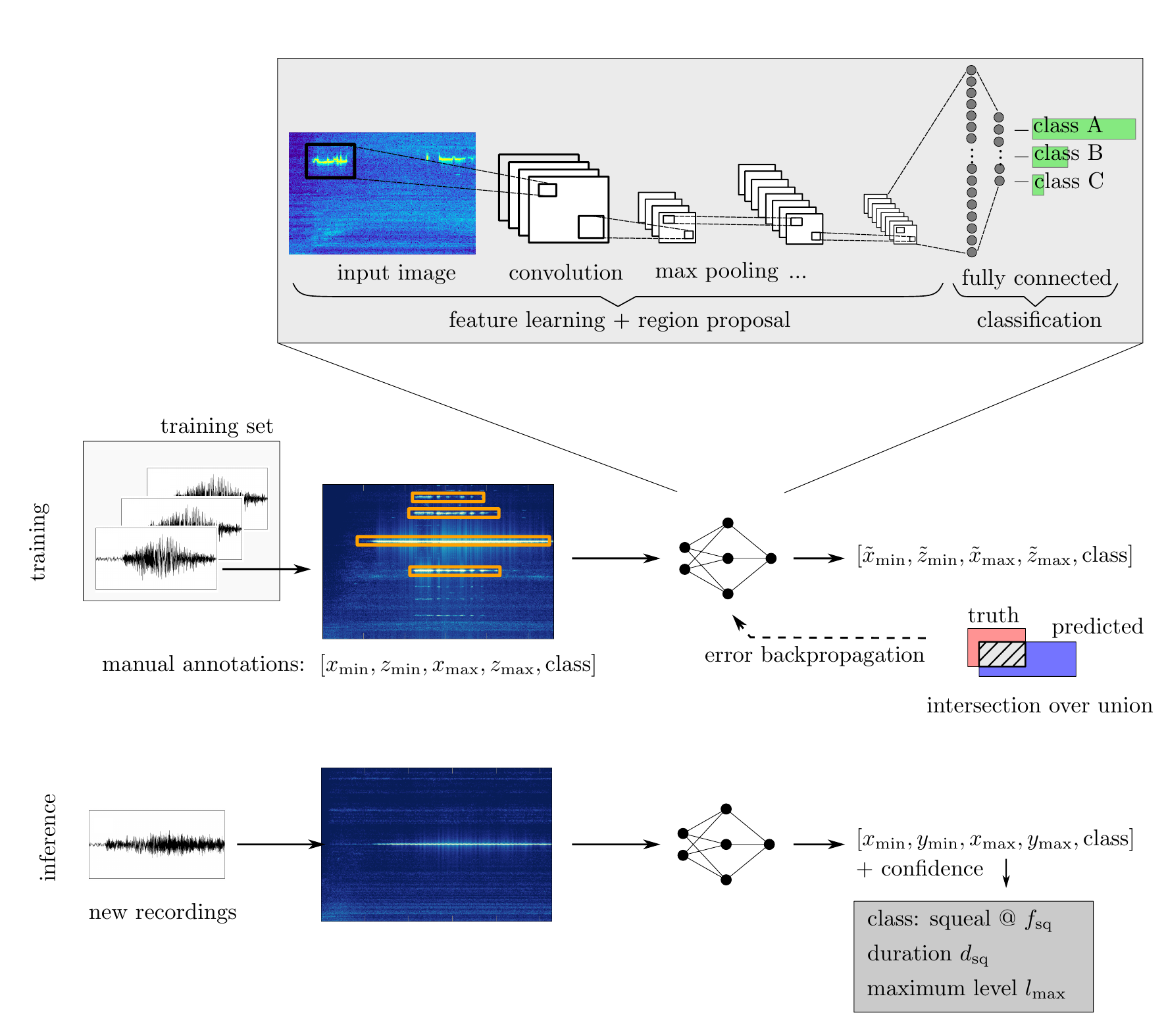}
\caption{Schematic illustration of the deep learning brake noise detector. The upper parts depicts the general structure of a CNN which is the backbone of the proposed method. During the training process, the parameters of the network are adjusted to learn the salient features of the objects fed into the network. The trained network can then be used to make predictions on new data, i.e. detect brake noise with a certain confidence. Objects are reported in terms of bounding box coordinates $x$ and $z$, which can be translated into frequency and time information related to the brake noise event}%
\label{fig:DLDetection}
\end{figure}

For the identification of noise classes in microphone recordings, we first transform the signals to two-dimensional representations using the short-time Fourier transform, i.e. by computing the spectrogram\footnote{The time-frequency precision of the DL detector can be controlled by the format of the spectrogram: the higher the resolution of the spectrogram used as input to the detector, the more precise the detection can be}. Then, object detection methods are applied to locate brake noises in the spectrogram. Figure~\ref{fig:DLDetection} illustrates this general approach schematically.  

Several thousand sound recordings acquired by microphone are available for training the novel detector. The data labeling (drawing bounding boxes and assigning labels) is performed manually by visual inspection and listening to the recordings. Labels are not limited to a single class per image, such that multiple bounding boxes of the four different classes, see Section~\ref{sec:requirements_detector}, can be assigned to a single spectrogram when multiple NVH events occurred during the single brake operation. Only a fraction of the  available data exhibited audible brake noises. In total, 3,276 braking operations remained for training, testing and evaluation of the different detectors. A set, hereafter denoted as the reference set, of 290 representative signals (200 squeals , 50 wirebrush noises, 25 click sounds and 15 broad-band noise signals) was kept aside for the evaluation of the detectors. The training set consists of 2,387 images and the test set consists of 599 images, corresponding to a $80-20$ split. 200 quiet brake operations were added to the reference data set as negative control. Each recording in the reference data set is a single-class image, so that multiple events within a single image belong to the same class. This set-up is required for a consistent evaluation of the conventional squeal detector which will serve later on as the base line model. 

\subsubsection{Model configuration and training}

We employ the concept of transfer learning to build deep learning noise detectors for NVH applications. Two instances of pre-trained models of the Faster R-CNN (inception and ResNet architecture) and an R-FCN model (ResNet architecture) from the Python Tensorflow model zoo\footnote{\url{https://github.com/tensorflow/models/blob/master/research/object_detection/g3doc/detection_model_zoo.md}} are considered starting point for the training process. Each model is trained for $100,000$ epochs using a unit batch size and the learning rate is set to be $0.0003$ for all models. To increase the network's robustness against overfitting, data augmentation strategies are used to horizontally flip, to randomly crop, to randomly pad, and to randomly add black patches to the images. In the following, \emph{DL model 1} refers to the deep learning brake noise detector built on the Faster R-CNN inception architecture, \emph{DL model 2} refers to the Faster R-CNN resnet architecture, and \emph{DL model 3} refers to the R-FCN ResNet architecture.

The performance of the conventional and the deep learning detectors are assessed using the reference set as ground truth. For each image, there exist ground truth bounding box locations as well as corresponding class labels. To evaluate the correct bounding box location, the intersection over union (IoU) metric is utilized, see \ref{app:IoU}. The IoU measures the overlap of the predicted box with the ground truth box. Box locations are labeled as valid if the IoU exceeds the threshold value of $\mathrm{IoU}>0.75$. Given a valid box location, the object class needs to be correct as well. The predicted labels are compared to the ground truth to compute the number of true positives (TP), false positives (FP), false negatives (FN), and true negatives (TN). Various classification quality metrics are constructed from those values, such as accuracy, precision, recall, true negative rate (TNR), and the $F_1$ score

\begin{align}
\mathrm{accuracy} &= \frac{\mathrm{TP}}{\mathrm{TP}+\mathrm{FP}+ \mathrm{FN}+\mathrm{TN}} \\
\mathrm{precision} &= \frac{\mathrm{TP}}{\mathrm{TP}+\mathrm{FP}} \\
\mathrm{recall} &= \frac{\mathrm{TP}}{\mathrm{TP}+\mathrm{FN}} \\
\mathrm{TNR} &= \frac{\mathrm{TN}}{\mathrm{FP}+\mathrm{TN}} \\
F_1 &= 2 \cdot \frac{\mathrm{precision} \cdot \mathrm{recall}}{\mathrm{precision}+\mathrm{recall}} \quad . 	
\end{align}

 An overly attentive detector would have a high sensitivity (\textit{recall} or \textit{true positive rate}, i.e. the number of squeals that are correctly detected), but a low specificity (\textit{true negative rate} i.e. the number of recordings in which the detector correctly did not find a noise event). The $F_1$ score is used to balance the precision and recall, which can be useful for imbalanced class distributions with $F_1=1$ indicating a perfect classification. 

Using softmax activations at the output layers, the DL models return a confidence score $C$ for each class that ranges from zero to one. A threshold score of $C_{\mathrm{min}}$ is considered for reporting a predicted class label. The optimal choice of $C_{\mathrm{min}}$ for each class is discussed in \ref{app:confidenceLevelStudy}. The confidence score can also be used to rank the predictions of all examples. Then, the cumulative precision and recall can be computed for decreasing confidence values. The result is typically displayed in form of the precision-recall curve (PRC). A good detector exhibits high precision values even for increasing recall values, hence a decreasing number of false positives while the number of false negatives stays low. Practically, this means that also for lower confidence values the model still detects most of the positive objects without reporting too many false positives. The area under the curve (AUC) then measures the average precision (AP) for a single class of objects. For multi-class tasks the mean average precision (mAP) summarizes the overall performance of the classification model. \ref{app:prc} elaborates further on the PRC used in this work for object detection evaluation. As the spectral squeal detector does not return confidence values, a synthetic confidence scoring, see \ref{app:confidence_score_spectral} is introduced based on the length and the intensity of the detected squeal sound.

\subsubsection{Classification performance}
 To obtain the classification performance of the DL models, the object detection task is transformed into a multi-class image classification task by neglecting the bounding box locations and only considering the predicted classes, questioning whether there is at all a noise found in the recording. As the images of the reference data set are single-class images, we can assign a unique class to each image, and then compare ground truth and prediction based on the class label.  
 
\begin{table}[h]
\centering
\caption{Classification performance of all detectors evaluated on the reference data set containing 200 quiet brake operations, 200 squeals, 50 wirebrush sounds, 25 click sounds and 15 broad-banded artefacts. Quality metric are evaluated for the three deep learning (DL) models and the baseline spectral squeal detector. Accuracy values are not reported as they are highly biased towards the over-represented squeal class.}
\begin{small}
\begin{tabular}{llcccc ccccc} \toprule
classifier & category & TP & FP & FN & TN & recall & precision & $F_1$ & AP & mAP\\ \midrule
\addlinespace
\multirow{3}{*}{DL model 1} & squeal 		& 149	& 68	& 7	& 266	& 0.96	& 0.69	& 0.8	& 0.68 & \multirow{4}{*}{0.73} \\
													& wirebrush 	&  33	& 1	&  17	& 439	& 0.66	&  0.97	&  0.79	&  0.69	&\\
\multirow{1}{*}{\scriptsize{(Faster R-CNN incep)}}	& click 	&	23& 8	&  2	& 457	& 0.92	& 0.74	& 0.82	& 0.88&\\
												& artefact 			& 10	& 0	& 5	& 475	& 0.67	& 1.00	& 0.80	& 0.68&	\\ 
\midrule
\addlinespace
\multirow{3}{*}{DL model 2} & squeal 	& 149	& 76	& 7	& 258	&	0.96	& 0.66& 0.78& 0.72 & \multirow{4}{*}{0.80}\\
												& wirebrush 	& 38	& 3	&	12& 437& 0.76	& 0.93	& 0.84 & 0.78	&\\
\multirow{1}{*}{\scriptsize{(Faster R-CNN ResNet)}} & click & 22 	& 6& 3& 459& 0.88& 0.79& 0.83 &	0.87&\\
												& artefact 	& 12	& 0 	&3  & 475 & 0.8	&	1.00& 0.89 &0.81	&\\ 
\midrule
\addlinespace
\multirow{3}{*}{DL model 3} & squeal	& 146 & 76	& 10	& 258& 0.94 & 0.66& 0.77& 0.68 & \multirow{4}{*}{0.55}\\
												& wirebrush & 26 	&	1& 24& 439 & 0.52 & 0.96 & 0.68	& 0.57 &	\\
\multirow{1}{*}{\scriptsize{(R-FCN resnet)}}& click 	& 21 	& 0	& 4	& 465	& 0.84 & 1.0	& 0.91 & 0.85	&\\
												& artefact 	& 1	& 0	&	14 & 475& 0.07 & 1.0& 0.12	& 0.1	&\\ 
\midrule
\addlinespace
\multirow{4}{*}{spectral} & squeal 		& 156 & 177	& 0	& 157	& 1.00	& 0.47	& 0.64 & 0.63 & \multirow{4}{*}{$-$}\\
												& wirebrush 	& $-$& $-$	& $-$	& 440	& $-$	& 	$-$& $-$	& $-$ &	\\
												& click 			& $-$& $-$ 	& $-$	& 465	& $-$	& $-$	& $-$ 	& $-$	&\\
												& artefact & $-$ & $-$ & $-$ & 475 &  $-$	&  $-$& 	$-$& 	$-$ &\\ 
\addlinespace
\bottomrule
\end{tabular}
\end{small}
\label{tab:classification_result}
\end{table}
 Table~\ref{tab:classification_result} reports the results of different detectors by their confusion matrix entries per category. Even though the spectral detector is designed only for squeal noise, its performance in the three remaining classes is reported as well\footnote{Ideally, this detector would return zero true positives, zero false positives and 490 true negatives for the categories \textit{wirebrush}, \textit{click} and \textit{broad-band noise}}. It can be observed that the spectral detector has a very high number of false positives. Further analysis reveal that it detects squeal noise in brake operations which are either quiet or which show wirebrush noise.  Obviously, this characteristic crucially depends on the parameter settings of the detector, such as the minimal peak sharpness, and minimal sound pressure level, among others. However, for practical employment in research and design, those spectral detectors are typically designed to be over-sensitive in order to avoid false negatives and to not overlook possible NVH issues. As a result, the detector exhibits no false negatives, i.e. finds all squeal events, for the data set studied here. Considering the deep learning detectors, all models achieve good classification performances for all four categories. Generally, the number of false positives is high for the squeal category, whereas the false negatives rate is higher for the wirebrush category. The squeal classification performance is similar for all DL models and reach scores ranging from 0.77 to $F_1=0.8$. For the three other categories, models 1 and 2 outperform model 3 as indicated by the $F_1$ scores. Especially, model 3 exhibits a high false negatives rate for noise artefacts and wirebrush, both representing spatially extended objects in the spectrograms. Overall, the faster R-CNN ResNet architecture (model 2) performs best in this study with similar squeal classification ability as the conventional approach and high classification quality for the other object classes, thereby representing substantial progress in brake noise detection tasks. 

\begin{figure}[htb]
\centering
\includegraphics[width=1.0\columnwidth]{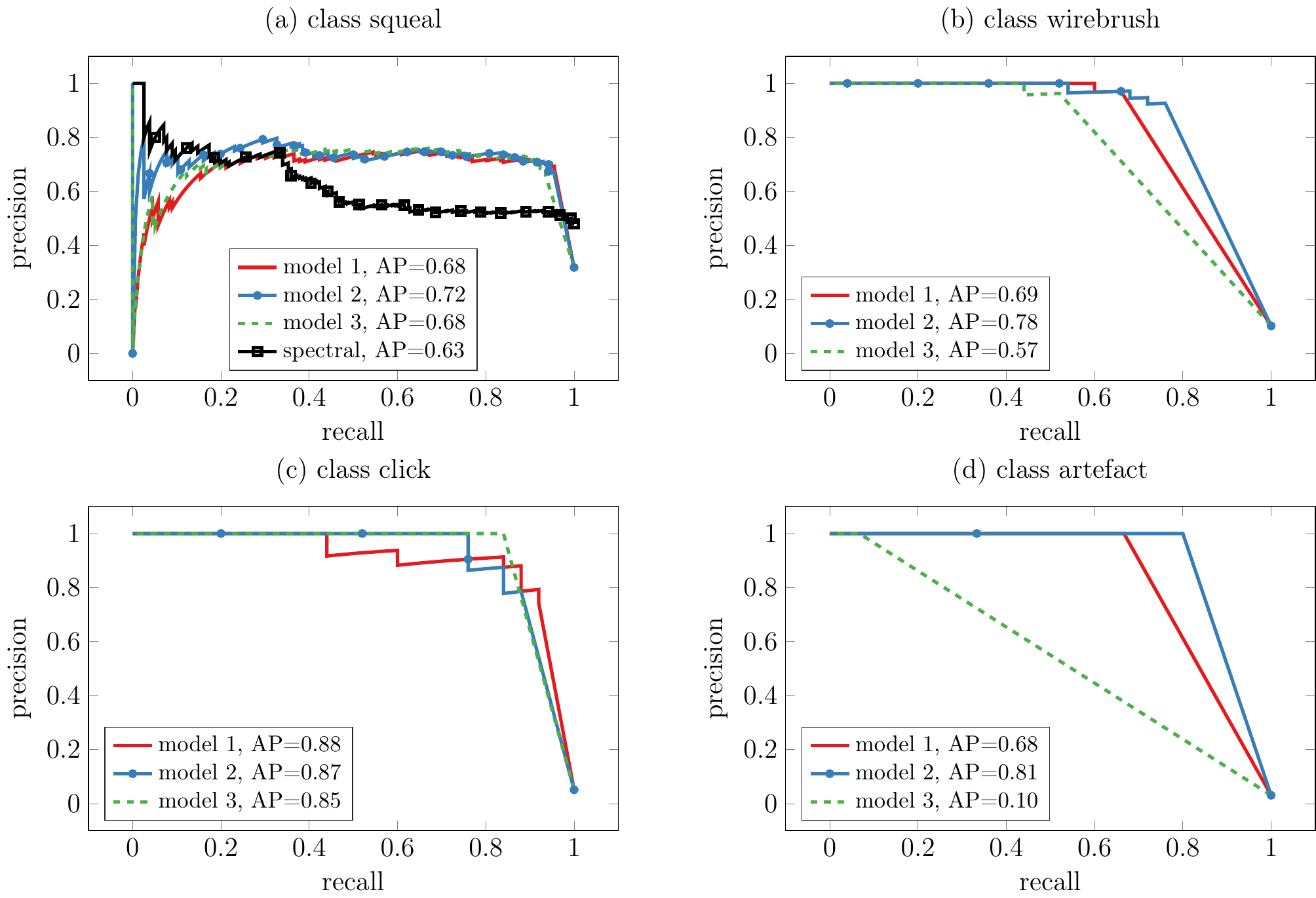}%
\caption{Precision-recall curves (PRC) for the deep learning detectors evaluated for the reference data set classification task. See \ref{app:prc} for more details}%
\label{fig:classification_PRC}
\end{figure}

Figure~\ref{fig:classification_PRC} depicts the precision-recall curves for the classification task. While the performance for squeal classification is rather similar for all DL models, the conventional squeal detector shows poorer performance for recall values larger than $0.3$. The superior overall performance of model 1 and model 2 becomes obvious for the wirebrush and artefact categories. Here, the PRC of model 3 drops significantly earlier for increasing recall values, thus representing poorer performance. Model 2 ($\mathrm{mAP}=0.8$) achieves the best overall score, followed by model 1 ($\mathrm{mAP}=0.73$) and model 3 ($\mathrm{mAP}=0.55$).  

Concluding, the deep learning based detectors exhibit improved capabilities compared to the spectral approach when considering the multi-class brake noise classification task. While spectral methods have to be designed for a low specificity resulting in many false positives, the deep learning detectors can be designed towards high sensitivity owing to multiple class labels.   

\subsubsection{Object detection performance}
In the object detection analysis, we consider also the time-frequency localization, i.e. not only the class label but also the correct bounding box location is taken into account. The object's location information allows to state \emph{when} and in \emph{which} frequency range the brake noise occurred. The minimum confidence levels for reporting a predicted object are equal to the ones chosen for the image classification task. Three variations of the minimum IoU threshold, i.e. the required bounding box overlap, are investigated. Table~\ref{tab:objectDetection_result} reports the average precision values per class and the resulting mean average precision per detector. The corresponding precision-recall curves can be found in \ref{app:objDetecPRC}.  

\begin{table}[h]
\centering
\caption{Object detection performance measured by average precision per class (AP) at IoU levels $50\%$, $75\%$ and $90\%$ and resulting mean average precision (mAP) for each deep learning classifier}
\begin{small}
\begin{tabular}{llcccc cccc} \toprule
classifier & category & AP$_{0.50}$ & AP$_{0.75}$ & AP$_{0.90}$ & mAP$_{0.50}$ & mAP$_{0.75}$ & mAP$_{0.90}$ \\ \midrule
\addlinespace
\multirow{3}{*}{ML model 1} & squeal 	& 0.66 & 0.66	&	0.60	& \multirow{4}{*}{0.66} & \multirow{4}{*}{0.65} & \multirow{4}{*}{0.59}\\
												& wirebrush  	& 0.57	& 0.54	&		0.36 	\\
\multirow{1}{*}{\scriptsize{(faster R-CNN inception)}}	& click  	& 0.74& 0.74	&	0.72				\\
												& artefact 		& 0.67	& 0.67	&	0.67 \\ 
\midrule
\addlinespace
\multirow{3}{*}{ML model 2} & squeal &0.68 & 0.68 &	0.62& \multirow{4}{*}{0.69} & \multirow{4}{*}{0.67} & \multirow{4}{*}{0.55}\\
												& wirebrush  	& 0.61& 0.51&	0.19 	\\
\multirow{1}{*}{\scriptsize{(faster R-CNN ResNet)}} & click & 0.80 & 0.80 &	0.71				\\
												& artefact	& 0.69& 0.67 &	0.55 	\\ 
\midrule
\addlinespace
\multirow{3}{*}{ML model 3} & squeal & 0.66 & 0.65&	0.60& \multirow{4}{*}{0.43} & \multirow{4}{*}{0.39} & \multirow{4}{*}{0.34}\\
												& wirebrush 	 	& 0.44& 0.30&	0.16 	\\
\multirow{1}{*}{\scriptsize{(R-FCN resnet)}}& click  	& 0.55 & 0.55 &	0.55	\\
												& artefact  	& 0.07& 0.07&	0.07	\\ 
\addlinespace
\bottomrule
\end{tabular}
\end{small}

\label{tab:objectDetection_result}
\end{table}

Overall, and similarly to the classification task, model 2 performs best. For the squeal category this detector shows slightly higher AP values than models 1 and 3. In the other categories, even higher quality metrics are observed. Model 3 fails to detect artefacts correctly, and also shows poorer performance for the wirebrush and click class. Model 1 exhibits a good overall performance in all categories, but cannot reach model 2. However, these detection quality metrics must be considered with care: the squeal and click objects are rather slim objects, while wirebrush and artefacts may cover substantial area within a spectrogram. Hence, the imposed IoU thresholds are much stricter for the slim objects than for the spatially extended objects which may create a bias. Naturally, the increase of the IoU threshold reduces the detection metrics: the higher the required bounding box overlap between ground truth and prediction is, the lower the number of true positives is. In this study, a minimal IoU of $0.75$ would be a good choice for obtaining high precision in the bounding box locations without losing too many correctly detected objects. For actual use cases in a data science process, the IoU and the confidence limit $C$ need to be tuned to meet specific requirements in terms of precision and recall.  

\begin{figure}[!h]
\centering
\includegraphics[width=0.99\columnwidth]{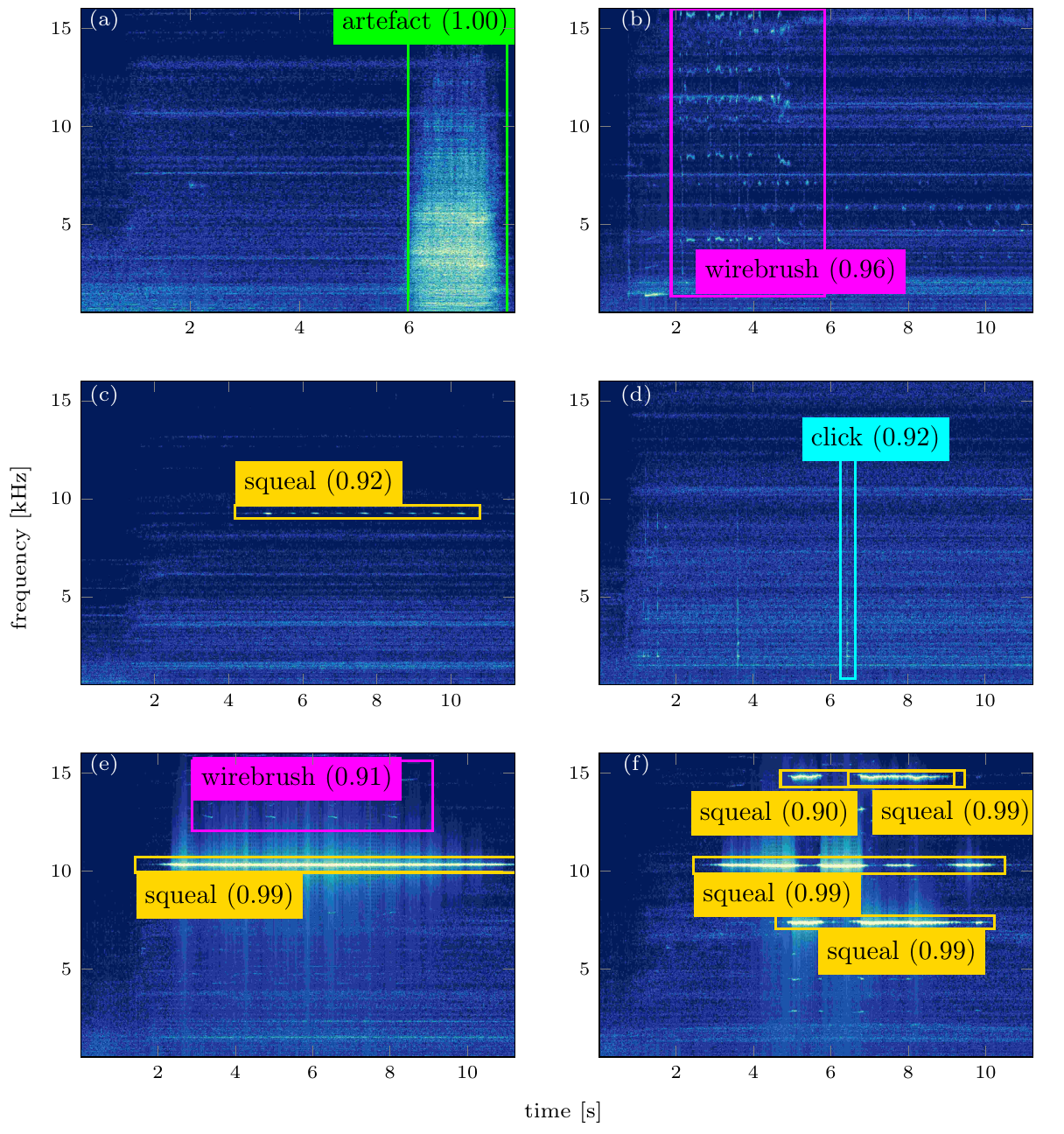}%
\caption{Results of the deep learning brake sound detector (model 1) applied to six microphone recordings. The detector can handle single class images, cf. (a-d), multiple events per image, cf. (f), and multi-class images, such as (e). The low intensity squeal in (c) is successfully identified, but the first click sound in (d) at $3.8$\,s is missed, therefore representing a false negative. The confidence scores are given in brackets}%
\label{fig:detector_examples}%
\end{figure}

Figure~\ref{fig:detector_examples} depicts a selection of qualitatively different brake sounds and the detection results obtained using model 1. Most of the bounding boxes are predicted at the correct locations and all class labels are correct. The model is capable of detecting single and multiple events in the spectrograms and returns high confidence scores. For the squeal category, it can handle temporal gaps, see case (c), for a single squeal frequency as well as spectral gaps, i.e. multiple co-existing squeal events separated in the frequency direction such as in (f). Multi-class predictions are also successful, see (e). In a last step the bounding box coordinates can be reported in terms of temporal, i.e. duration, and frequency information. As the bounding box may not be perfectly centered around the true frequency, the conventional squeal detector has a higher precision in reporting the frequency value.

To be employed in a real testing environment, the detection models should to be ideally able to analyze data in the stream and possibly on notebooks. To facilitate this in the future, all models were trained on a GPU-supported laptop \footnote{specifications used here: Intel i$7-8750$ CPU 6x $2.20$\,GHz, $16$\,GB RAM, NVIDIA GeForce GTX 1050Ti $4$\,GB} using the Python-based deep learning framework TensorFlow. While training time for $10^6$ epochs took approximately a complete day for each model, the time to conduct the inference is rather short: On average, $0.17$\,s per recording for model 1, $0.45$\,s per recording for model 2 and $0.24$\,s per recording for model 3 were measured. This highlights that once the model is trained a very fast response time can be expected. 

Future research activities are possible in two major aspects of the proposed methodology. First, other detection models may be tested for their applicability and performance. Here, we think of other deep learning object detection architectures as well as conceptually different approaches such as classical CV methods. For edge detection, the Canny edge detection filter \cite{Canny.1986} could be applicable for squeal and click sound localization. Second, fine-tuning of the deep learning models may leverage their full potential and increase the detection quality even more. It could also be of advantage to combine methods used in nonlinear time series analysis as additional metrics to especially detect nonlinearities and to reveal the physical nature of the signal in the sense of knowledge discovery \cite{Stender.2019c,Stender.2019f}. Finally, a combination of spectral and DL detectors is possible to cross-validate the detection of a single method and to combine advantages of both methods. The second part of this work will use the deep learning squeal detector for labeling large amounts of data from brake system NVH testing.

\section{Part 2: Brake squeal prediction}

With part 1 providing labels for large data sets acquired during experimental NVH testing on brake dynamometers, this part aims to train deep neural networks for predicting the instantaneous vibration behavior of the system from its loads. The objective of this work is to study whether there is a deterministic relationship between the loading conditions of the brake system and its behavior in terms of friction-induced noise. If the system's noise performance can be predicted from those inputs, the identification of instability regimes is possible, which in turn allows forecasting the onset of squeal for a given loading scenario. Given four experimental data sets and using recurrent neural networks, we aim to answer two fundamental questions:
\begin{enumerate}
	\item Is brake squeal predictable from the recorded operational conditions using deep learning for a single brake system?
	\item If squeal is predictable, is there an underlying mechanism that is immanent in multiple brake systems irrespective of geometries or configurations?
\end{enumerate}

This part is organized as follows: first, the parameter-dependent brake system behavior is re-visited before time series classification techniques and recurrent neural networks are introduced. Then, data pre-processing steps and a hyperparameter study are discussed to find an appropriate network configuration for the given classification task. Classifiers are trained on four data sets with two different objectives: first to predict if a given set of loads will cause squeal, and second to predict at which time instance the excitation of the squeal-vibrations takes place. Lastly, models trained on one system are evaluated on a different brake system to investigate if such networks can generalize. 

\subsection{Parameter dependence and brake noise behavior}

Besides model parameter uncertainties \cite{Zhang.2016}, and multi-stability scenarios \cite{Stender.2020b}, the huge bifurcation parameter space is one of the major reasons for the poor squeal prediction quality of numerical simulations even today \cite{Butlin.2010}. Therefor, the inherently transient \cite{Pilipchuk.2015} and multivariate input (loads) - output (vibration response) behavior of brake systems is studied hereafter. Physical measurements of operational loads of the brake system are considered as proxies that drive main stiffness and damping variations \cite{Sinou.2007, Sinou.2007b} in the system. Formally, one may express this system understanding as parametric velocity-dependent $\mathbf{P}$ and displacement-dependent $\mathbf{Q}$ model terms, strongly nonlinear forces $\mathbf{f}_{\mathrm{nl}}$ from the friction interface, from joints and from geometric constraints, as well as external forces $\mathbf{f}_{\mathrm{ext}}$. The understanding can be reduced to a parametric formulation of the classical equations of motion

\begin{equation}
\mathbf{M}\left(\mathbf{\Xi} \right) \ddot{\mathbf{x}} + \mathbf{P}\left( \mathbf{\Xi} \right) \dot{\mathbf{x}} + \mathbf{Q}\left( \mathbf{\Xi} \right) \mathbf{x} + \mathbf{f}_{\mathrm{nl}} \left(\mathbf{x}, \dot{\mathbf{x}}, \mathbf{\Xi} \right) = \mathbf{f}_{\mathrm{ext}} \left( \mathbf{\Xi} , t\right) , \quad \mathbf{\Xi} = f\left(\mathbf{x}, \dot{\mathbf{x}}, t \right) \in \mathbb{R}^{\tilde{m}}
\label{eq:eom}
\end{equation}

that govern the system dynamics. Bifurcations can occur as members of the parameter vector $\mathbf{\Xi}$ change, for example temperature or wear \cite{AWREJCEWICZ.2013}, which in turn depend on system states. Here, the $m$ measured loads are considered to constitute the multivariate and time-dependent parameter vector $\mathbf{\Xi} \in \mathbb{R}^{m}$. For a complete system description, the parameter vector would have to contain a, probably, uncountable number of $\tilde{m}$ quantities that can act as a bifurcation parameter for the system dynamics, and the full multi-physic system description would take the form of partial differential equations. Hence, one of the major objectives is to find out whether the $m$ loads measured during the testing campaign represent dominant members of the parameter vector $\mathbf{\Xi}$ and therefore allow for instability prediction. Thinking of brake squeal as a parameter-driven behavior is motivated by the architecture of neural networks whose primary task is the learning of a function approximation between input and output values. We still consider self-excitation through flutter or a falling friction slope, and not external forcing, as the instability mechanisms of brake squeal. However, to initiate instability, parametric changes and subsequent bifurcations in the system are required. Those changes of parameters are encoded in the load sequences taken as inputs to the network, and the vibrational response represents the output of the network. 
\begin{figure}[htb]
\includegraphics[width=1.0\textwidth]{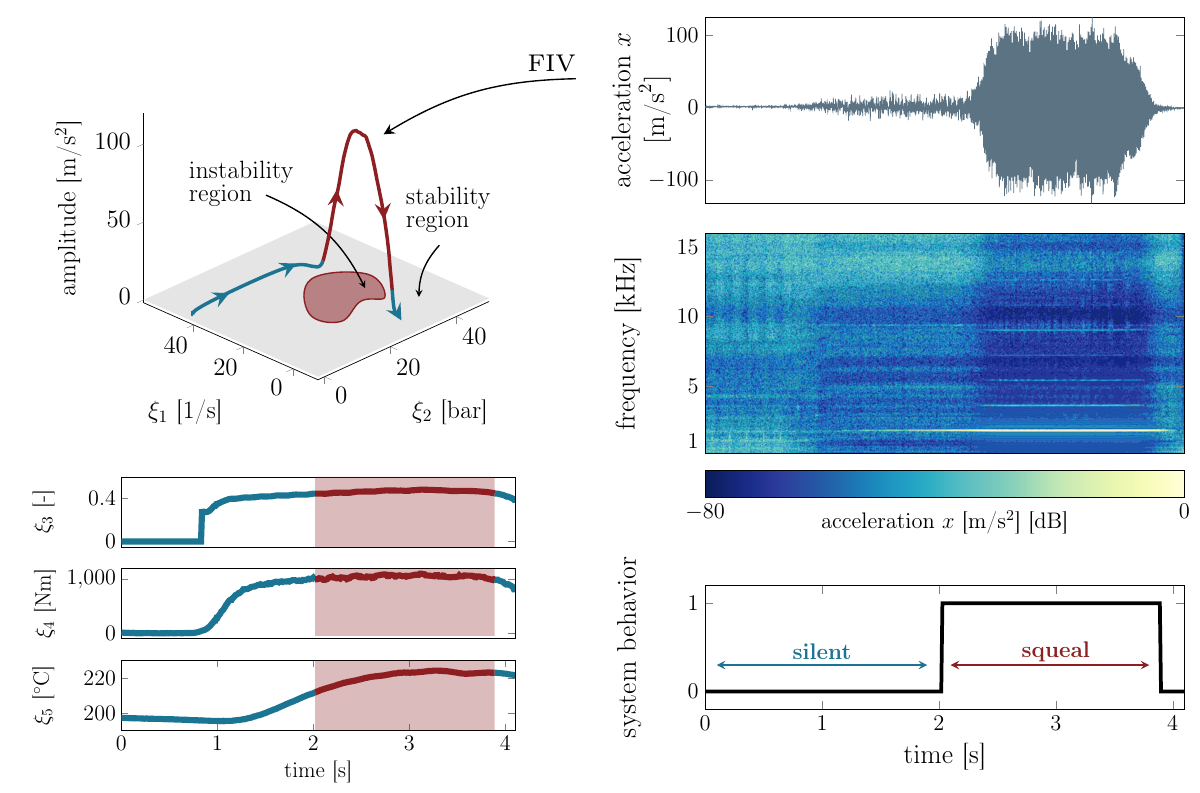}
\caption[]{Brake squeal as a dynamic instability induced by parameter variations (illustration inspired by \cite{Sapsis.2018}). During braking, loads and operational conditions, such as $\xi_1$ (rotational velocity $\Omega$), $\xi_2$ (pressure $p$), $\xi_3$ (friction coefficient $\mu$), $\xi_4$ (brake torque $M$) and $\xi_5$ (disk surface temperature $T_{\mathrm{rot}}$), change. The mechanical system exhibits instability regions at different combinations of those loads as schematically indicated in the $\xi_1 - \xi_2$ parameter plane. High-intensity, noise generating friction-induced vibrations (FIV) arise once the system enters an unstable parameter regime. While the loads change rather slowly, vibrations grow instantaneously and show a rich and transient spectral content. For the time series classification task, the loads $\xi_i\left(t\right)$ are considered as input variables and the system response represents the target variable. For that, the vibration measurement is encoded into the time-dependent state of \textit{silent} (0) and \textit{squealing} (1) behavior}
\label{fig:instability}
\end{figure}

Figure~\ref{fig:instability} illustrates loads and the vibrational response recorded during a single braking on a dynamometer. There may exist multiple regions of instability in the parameter space $\mathbf{\Xi} = \left[ \xi_1, \dots, \xi_m \right]$ with possibly complicated boundaries. Even if the loads $\xi_i$  would represent the full bifurcation parameter space, manual discovery of the multi-dimensional and time-dependent patterns leading to instability would be impractical, if not impossible. Hence, recurrent neural networks are employed to map the time-variant loading parameters to the squealing behavior, which represents a time series classification task. 

\subsection{Time series classification}
The format of the experimentally acquired data, i.e. loads and vibrations, takes the form of time series. Sampling a continuous quantity $s\left(t \right)$ at time instants $t_i$ results in an univariate time series $\mathbf{s}$ of length $n_{\mathrm{t}}$ 
\begin{equation}
\mathbf{s} = \left( s(t_1), s(t_2) \dots, s(t_{n_{\mathrm{t}}}) \right), \quad n_{\mathrm{t}}\in \mathbb{Z}^{+},
\label{eq:timeSeries}
\end{equation}
where a uniform sampling $t_{i+1}=t_i + \Delta t$ is assumed. Multivariate time series $\mathbf{S}(t) = \left[ \mathbf{s}_1, \mathbf{s}_2, \dots \mathbf{s}_m\right]$ store multiple time series in a contemporaneous fashion. Most importantly, the sequential order of the time series entries carries information about derivatives and history, and must hence be taken into account in the analysis. Time series classification (TSC) denotes the task of assigning a class label to a time series. Bagnall et al. \cite{Bagnall.2017} propose to cluster TSC approaches into three main categories. (1) \textit{Distance-based classifiers} \cite{Wu.2018a}, also referred to as \textit{instance-based classifiers}, measure similarity between different time series in terms of Euclidean or other distances and then assign class labels. (2)  \textit{Feature-based classifiers} \cite{Christ.2018} transform the dynamic time series into a set of static values that describe certain properties of the sequence, e.g. the mean, variance or others. To derive discriminating features, expert domain knowledge is required. Those static features can then be fed to any conventional machine learning algorithm to assign a class label. (3) \textit{Direct approaches} \cite{Langkvist.2014} learn representative high-level features themselves and do not rely on manually hard-coded feature extraction recipes. Hence, direct approaches receive the raw time series as input and output the class label. Direct approaches do not require a-priori domain expert knowledge to extract discriminating features and are generally able to learn complex temporal patterns. As a downside, larger amounts of data are required during the training phase owing to the increased model complexity. The architecture of direct approaches can be constituted of convolutional neural networks (CNN) \cite{Yang.2015, Kiranyaz.2016}, recurrent neural networks (RNN) \cite{Husken.2003} or others \cite{Swapna.2018}. In this work we propose to use a variation of RNNs, so-called Long-Short-Term Memory networks (LSTM) \cite{Hochreiter.1997}, for the prediction of brake squeal. In the data sciences, the term \emph{prediction} refers to computing the output of a trained model for a given input. In the context of brake noise, and with regards to the actual use case in this work, we will utilize the term \emph{prediction} in the sense of a virtual twin forecasting the brake noise behavior for a set of loading scenarios.

\subsection{Recurrent neural networks}
Recurrent neural networks (RNN) take sequential inputs into account by creating a time dependent memory state $\mathbf{h}(t_i)$ that carries information about previous states in time. At time $t_i$ the memory state is evaluated with respect to its previous state
\begin{equation}
	\mathbf{h}(t_i) = \sigma\left(\mathbf{h}(t_{i-1}), \mathbf{S}(t_i)\right)
	\label{eq:RNN_memoryState}
\end{equation}
where we assume a multivariate time-series input $\mathbf{S}(t_i)$ and a nonlinear and differentiable activation function $\sigma(\cdot)$. With this memory state RNNs are capable of taking the history of the time series into account when evaluating the current value at time $t_i$. Processing the input $\mathbf{S}(t_i)$ results in a multivariate output of the RNN layer, which again might be used as input to the next layer. Multiple RNN layers can be stacked to generate a deep network architecture. To surpass the \textit{vanishing gradients} effect \cite{Hochreiter.1997,Pascanu.2013} of classical RNNs, and to enable longer memory capacity of the network, long short-term memory (LSTM) networks \cite{Hochreiter.1997} are employed in this work. A common application of RNNs and their modifications has been the analysis of human centered sequential data as used in speech and handwriting recognition \cite{Graves.2013, Sak.2014, Greff.2017, Graves.2008}. Furthermore, RNNs are used for medical application, e.g. for analyzing ECG data \cite{Silipo.1998}. In the mechanical engineering disciplines, application cases have been reported for predictive maintenance \cite{Zheng.2017}, system health monitoring \cite{Khan.2018, Zhao.2019} and fault diagnosis for rotating machinery \cite{Liu.2018}. 

\subsection{Data preparation}

We study four different data sets $A$, $B$, $C$, $D$ in this work. All data sets stem from commercial testing similar to the SAE-J2521 procedure \cite{SAEJ.2521} on an NVH dynamometer. Squeals are detected in the microphone recording for a minimum squeal duration of $d_{\mathrm{sq}}=0.5$\,s and a minimum sound pressure level of $l_{\mathrm{sq}}=55$\,dB(A). Data sets $A$ and $B$ stem from the same family, i.e. the brake systems are similar in terms of geometry and performance. Data sets $C$ and $D$ stem from two brake systems which are similar to each other but significantly different to $A$ and $B$. Hence, rather similar dynamic behavior can be expected for members of the same family, while large differences between families will not be surprising. However, due to the elusive and highly sensitive character of brake squeal and multiple sensitivities, such assumptions must be confirmed by the analysis.

The data sets differ with respect to the number of brakings and the number of squealing brakings: $A$ (1206 brakings, 487 of which squealing), $B$ (1206 / 227), $C$ (1206 / 347), $D$ (1889 / 237). Designed to quietly dissipate brake energy via heat, a brake system's operations typically exhibit only a small number of squeal noises. For this study particularly unstable brake systems were selected to obtain more squeal samples. Still, \emph{class imbalance}, i.e. the under-representation of squealing brake operations, can be observed. For each braking, the $m=8$ loading conditions of disk velocity $\Omega$, brake line pressure $p$, brake torque $M$, friction coefficient $\mu$, disk surface temperature $T_{\mathrm{rot}}$, brake fluid temperature $T_{\mathrm{fluid}}$, ambient temperature $T_{\mathrm{amb}}$ and relative humidity $H$ are available as time sequences. Furthermore, for each time step the binary vibration label \textit{squeal} / \textit{quiet} (1/0) is available. The lengths of the time records vary between $2$\,s and $10.5$\,s, see \ref{app:data_char}. As the training and performance of LSTM networks can be significantly facilitated through equal length inputs, a sliding window pre-processing step is introduced. The input and output sequences are segmented into windows of $w$ samples. The shift parameter $h$ denotes the number of samples by which the windows shift and $(w-h)$ provides the window overlap. Observations at the end of sequences are zero-padded to full length $w$ if they are longer than half the window size. Shorter sequence remnants are omitted. Depending on the choice of $w$ and $h$, the number of observations available for training can be significantly increased. Even if LSTMs are designed to learn long-time correlations, it is physically unknown which time history is required to predict the onset of squeal at the current time instance. The instability may be rooted in an instantaneous conditions, but can also be initiated through load history effects that require the consideration of longer sequences. Hence, the sliding window parameters are treated similarly to the hyperparameters of the network. Hyperparameters refer to the structure-related meta-parameters of a neural network, for example to the number of layers. A hyperparameter study aims at finding an optimal configuration of the network. 

\subsection{Design of virtual NVH twins}
\begin{figure}[ht]
\centering
\includegraphics[width=0.98\textwidth]{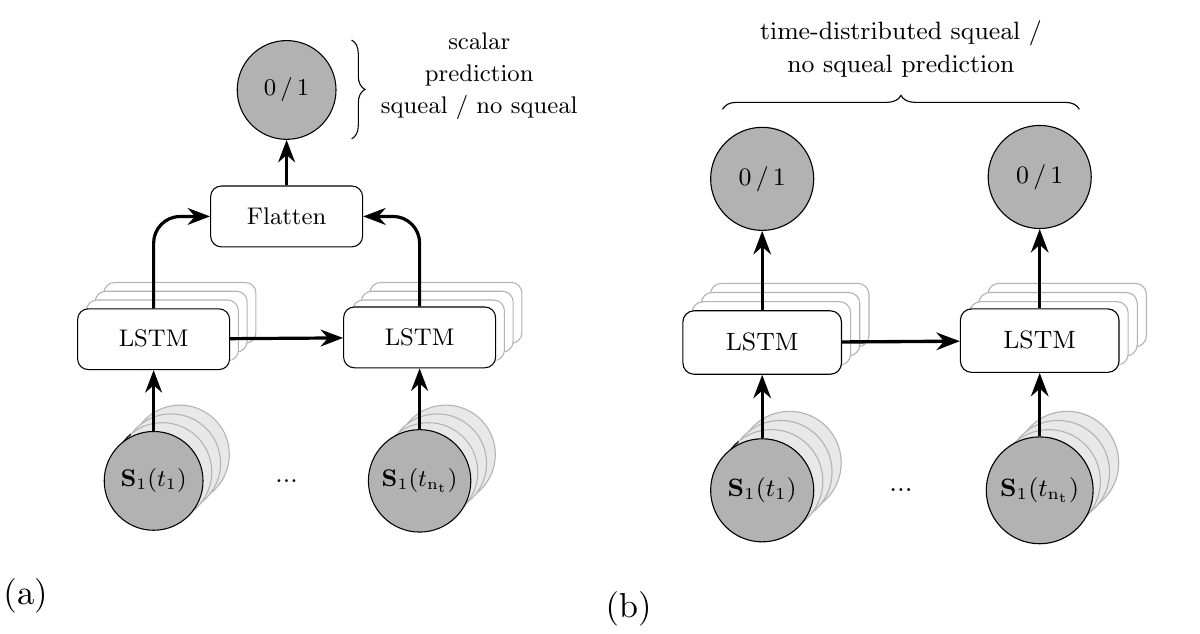}
\caption{Two modes of brake squeal prediction: (a) 'sequence-to-scalar' prediction of a single label for an observation of loading sequences and (b) 'sequence-to-sequence' prediction of a binary label per time step. The dimensions of the inputs, layers and outputs are given in Table~\ref{tab:model_specs}}
\label{fig:OutputDifferences}
\end{figure}

Generally, two different digital twin architectures are studied, as schematised in Figure~\ref{fig:OutputDifferences}. First, the learning task is set up as a scalar binary classification task: for the multivariate input sequences, the model output is a boolean value (\emph{true} / \emph{false}) indicating \emph{if} squeal occurs during the braking. This set-up is referred to as \emph{sequence-to-scalar} in the following. Second, the model output is changed to a boolean sequence of the same length as the inputs. Hence, the output indicates squeal at every time instant. This \emph{sequence-to-sequence} setting also allows to predict \emph{when} a squeal occurs. This set-up is somewhat in analogy with the first part of this work, when labels were assigned to the images in the first step, and the location of objects was specified in the second step. 

To account for the severe class imbalance, that is the under-representation of squealing brakings as opposed to the over-representation of quiet brakings, the classification performance is measured by Matthews correlation coefficient (MCC) \cite{Matthews.1975}. The MCC 
\begin{equation}
\text{MCC} = \frac{\text{TP} \cdot \text{TN}-\text{FP}\cdot \text{FN}}{\sqrt{\left(\text{TP}+\text{FP} \right) \left(\text{TP}+\text{FN} \right) \left( \text{TN} + \text{FP}\right) \left( \text{TN}+\text{FN}\right)}}
\label{eq:MCC}
\end{equation}
ranges from $(-1)$ to $(1)$, where MCC$=-1$ indicates complete disagreement and MCC$=1$ indicates perfect classification. The uneven class distributions are explicitly considered in the MCC, such that this quality metric is not biased towards the over-represented class. A coarse hyperparameter grid search is carried out for finding appropriate LSTM model configurations, see \ref{app:lstm_config}. Adding a second LSTM layer did not substantially improve the score, so a single LSTM layer was chosen. The sequence pre-processing parameters $w$, $h$ and the network optimizer choice had only a secondary impact. Generally, the hyperparameter study is performed to obtain a first overview on the required model complexity. It is very likely that extensive hyperparameter optimization will further improve the classification scores reported hereafter. 

\subsection{Squeal prediction using a sequence-to-scalar classifier}
\label{sec:seq2scal}

The final classifiers for each data set are trained using a stratified, i.e. class distribution preserving, $70$-$30$ training-validation split.  Figure~\ref{fig:traininghistory_seq2scal} in \ref{app:lstm_training_history} displays the MCC curves for the training and validation set to demonstrate that the models are not overfitting. To obtain more representative results and reduce the potential bias caused by the data splitting, ten individual models are trained for ten repeated splits of the data sets. For these ten models, the average validation MCC and its standard deviation are reported in Table~\ref{tab:results_seq2scal}.  

\begin{table}[h]
\centering
\caption{Sequence-to-scalar classification result for all data sets. Ten models were fitted per data set, and the average and standard deviation of the validation scores are listed}
\begin{tabular}{c c c c}
\toprule
data set & training set size & validation set size & MCC mean$\pm$ std \\
\midrule
\addlinespace 
$A$ & 5354 & 2295 & $0.78 \pm 0.02$ \\
$B$ & 5325 & 2283 & $0.72 \pm 0.03$ \\
$C$ & 5462 & 2342 & $0.65 \pm 0.02$ \\
$D$ & 8614 & 3693 & $0.64 \pm 0.03$ \\
\bottomrule
\addlinespace
\end{tabular}

\label{tab:results_seq2scal}
\end{table}

As the model configuration was chosen to perform optimally for data set $A$, see \ref{app:lstm_config}, the highest classification scores are obtained for this data set. Nonetheless, the scores for the other data sets are well above MCC values of $0.60$, which indicates good classification performance. Moreover, the variation of the classification scores per data set is low, indicating non-bias towards  data splitting. Interestingly, the model performance reflects the brake system families: similar values are obtained for similar brake systems. As systems $C$ and $D$ are not similar to $A$ and $B$, lower scores are obtained for the former. Even if the data set $D$ is comprised of an extended testing matrix, i.e. more data samples are available, the model scores are the lowest. Hence, the instability conditions for systems $C$ and $D$ are either more complex, or the measurement channels carry less relevant pieces of information such that the classifiers exhibit lower prediction scores compared to systems $A$ and $B$. Yet, the validation scores clearly indicate the existence of deterministic instability patterns in the load-response behavior of all brake systems. To better understand the emergence of noise from an engineering perspective, more insights into the models are yet desired. As a first approach, the sequence-to-sequence classifiers will predict the point of vibration onset, which in turn allows to study instantaneous loading conditions. 

\subsection{Squeal prediction using a sequence-to-sequence classifier}
\label{sec:seq2seq}

\begin{figure}[ht]
\includegraphics[width=1.0\textwidth]{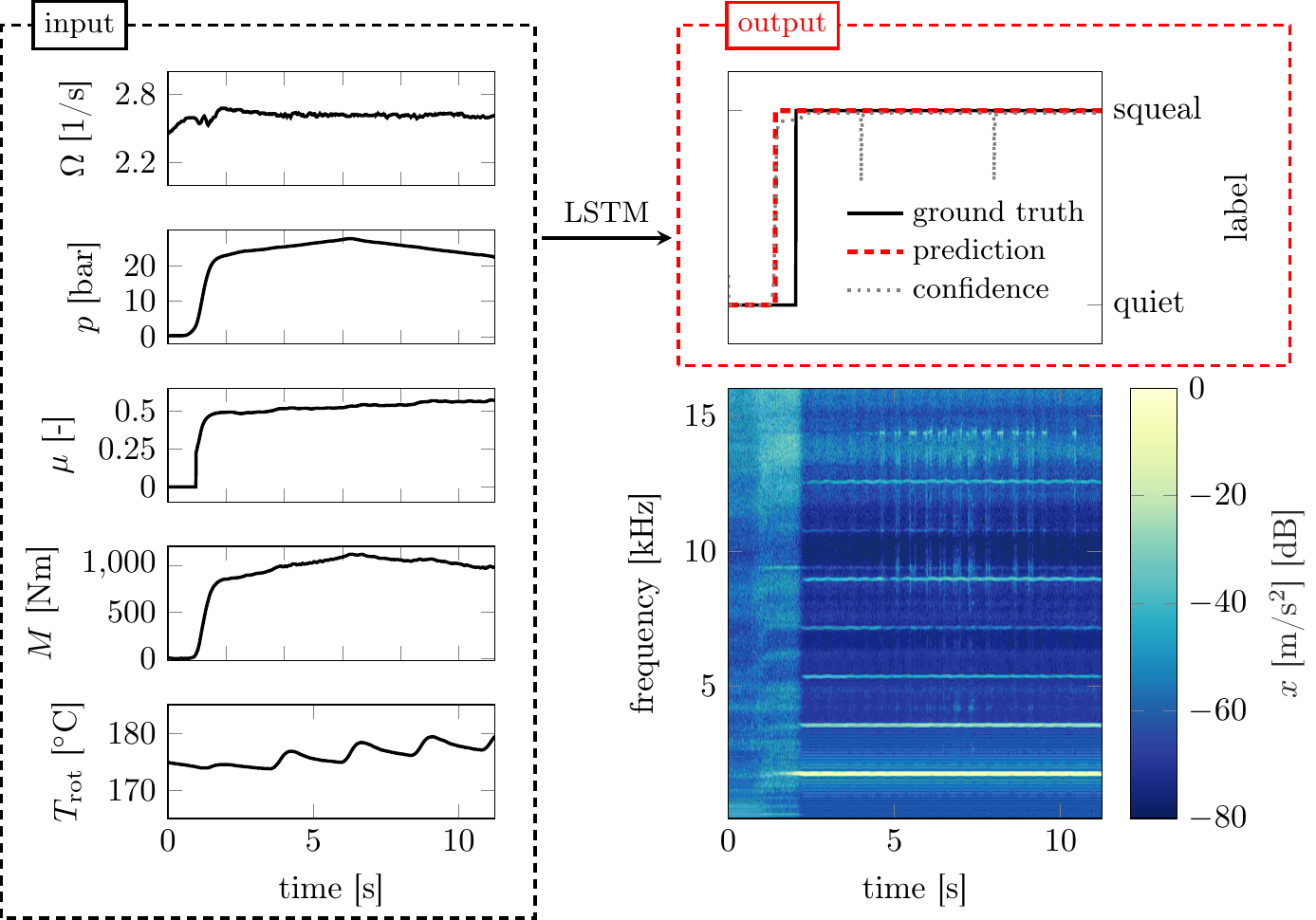}
\caption{Output of the sequence-to-sequence classification model for a validation sample: left panel displays the loading signals used as input sequences for the model. The top right panel depicts the model prediction as well as the model's confidence value given by the deviation of the dotted line from the dashed line. The lower right panel shows the spectrogram of the accelerations measured during braking. Overall, the network predicts the onset and duration of the squeal very well}
\label{fig:results_seq2seq}
\end{figure}

The second classification task involves more complex models that output a sequence of binary labels indicating the occurrence of squeal at each time point $t_i$. The hyperparameter study, see \ref{app:lstm_config}, showed better model performance for input sequences with a length of $w=400$ samples. As longer sequence lengths are required compared to the scalar classification case\footnote{here, $w$ was set to $200$ samples}, one can conclude that longer temporal correlations are playing a role in this squeal prediction set-up. 
\begin{table}[h]
\centering
\caption{Sequence-to-sequence classification result for all data sets using a fixed network configuration. $10$ models were fitted per data set, and the average and standard deviation of the validation scores are listed}
\begin{tabular}{c c c c}
\toprule
data set & training set size & validation set size & MCC mean$\pm$ std \\
\midrule
\addlinespace 
$A$ & 2827 & 1212 & $0.78 \pm 0.02$ \\
$B$ & 2785 & 1195 & $0.75 \pm 0.02$ \\
$C$ & 2893 & 1240  & $0.62 \pm 0.03$ \\
$D$ & 4536 & 1945 & $0.50 \pm 0.06$ \\
\bottomrule
\addlinespace
\end{tabular}

\label{tab:results_seq2seq}
\end{table}
Overall, classification scores mostly above $\mathrm{MCC}=0.5$ can be observed, as reported in Table~\ref{tab:results_seq2seq}. Again, the highest scores are obtained for data set $A$ and the deviations introduced by the individual data splits are small. Also, the two-class behavior between the first family of systems $A$ and $B$ and the second family of systems $C$ and $D$ is obvious. For the given model configuration, it seems to be easier for the networks to predict the onset and duration of squeal for the first two brake systems. Direct comparison of MCC values between the sequence-to-scalar and sequence-to-sequence classifiers of the MCC values is not possible: in the first case, a label is predicted per sequence, while in the second case a sequence of labels is predicted. Hence, if the scalar classifier correctly predicts squealing behavior per braking, high scores can be achieved. The sequence-to-sequence classifier needs to make correct predictions for each time step, which is a much stricter requirement. Hence, the classification scores reported in Table~\ref{tab:results_seq2seq} are related to a classifier that achieves a much more complex task than the scalar classifier, although the performance metrics seem to be on similar levels. Consider Figure~\ref{fig:results_seq2seq} as an example for the practical usage of a sequence-to-sequence classifier that can predict the onset of squeal. A braking from data set $A$ is presented that was not part of the training process. Predictions are made for the loading conditions sliced into segments of $400$ samples. The fluid and ambient air temperatures as well as the relative air humidity were supplied as additional inputs, but as they are almost constant, they are not displayed here. For this type of \textit{drag} braking, the disk rotation is kept constant and the pressure is varied. High-intensity squeal at a frequency of $1.8$\,kHz is excited at $t\approx 2.0$\,s and is sustained until the end. The classifier predicts squeal to set in at $t=1.8$\,s and to last to the end of the braking. Hence, the classifiers seems to have learned patterns and instability conditions that cause squeal from the training data. Those conditions are met in the time span from $t=1.8$\,s on, such that the model correctly predicts the squeal behavior here. The classification score obtained for this braking is $\mathrm{MCC}=0.85$, which illustrates how strict the MCC penalizes the false predictions in the region of squeal onset, i.e. the deviation of the model prediction from the ground truth.

The confidence score indicated by the network output from the sigmoid activation is high throughout the complete braking duration\footnote{In the graph, the vertical elevation of the \emph{quiet} marker relates to $0\%$ squeal propensity, and the elevation of \emph{squeal} relates to $100\%$ squeal propensity}. The slicing of the $11$\,s braking duration into three segments becomes visible through reduced confidence score at the very beginning of each sequence. In the first epochs of a sequence, the recurrent network has only a short time history available for making predictions, such that the classification confidence is smaller. 

Overall, it has been shown that squeal is predictable for individual brake systems from eight loading conditions using recurrent neural networks trained on rather small data sets. This is a very promising result for both the scientific and the commercial communities involved with disk brake squeal. First, neural network-based analysis and data driven approaches may help to understand the phenomenon better. Second, such models can constitute digital twins for the NVH behavior of brake systems during the engineering design phase, which can help to reduce the amount of hardware testing through faster, less cost-intensive and eco-friendlier virtual development. 

\subsection{Cross-evaluation}
The ultimate objective of most research activities in the field of brake noise is to find governing physical principles that explain the driving mechanisms for squeal and its parametric dependencies, i.e. stability boundaries. For example, the falling friction slope and the mode coupling instability represent such generic patterns of interest involving bifurcation parameters like the sliding velocity. To this end, we employ the concept of cross-evaluation to access how well a data-based classifier can predict the brake noise behavior of the present brake system even if the classifier was trained on data stemming from a different brake system. Such a scenario is natural for systems that are assembled by many components via mechanical joints: it has not only been shown that the joints' properties are highly variable but also that the joint-induced damping can be significantly larger than material damping \cite{Tiedemann.2015b,Grabner.2014, Brake.2018}. Hence, whenever a system is re-assembled or newly mounted onto a test rig, the dynamics may turn out to be fundamentally different to a previous test. However, as we do not supply dynamic system properties, e.g. modes, to the network, a successful cross-evaluation study would be in fact surprising. 

For the cross-evaluation, each classifier built on an individual data set in Sections \ref{sec:seq2scal} and \ref{sec:seq2seq} is used to predict the brake squeal behavior of the other three data sets: the classifier $\mathfrak{C}_{\mathrm{A}}$ built on data set $A$ is used to predict squeal in data sets $B$, $C$ and $D$, and like-wise for the other classifiers, without further training. Hence, the classifiers have not seen any data from the different brake systems before. If a classifier built on data set $i$ performs well on data set $j$, we can conclude that the model generalizes well for data sets $i$ and $j$. Hence, the underlying instability regions of brake systems $i$ and $j$ are either very similar, or the classifier may have learned a general physical mechanism that in fact governs brake squeal irrespective of the specific brake system realization. On the contrary, if poor prediction results are obtained, the instability regimes of brake systems $i$ and $j$ are either very different, or there is not a governing physical instability mechanism that can be read from the available data using the chosen models. As the pairwise data sets $A$, $B$ and $C$, $D$ stem from similar brake systems, those hypothesis can be tested on the data at hand. All data sets are prepared in the same manner, and the classifiers $\mathfrak{C}_i$ share the same network architecture given in \ref{app:lstm_config}.

The results of this cross-evaluation experiment for the sequence-to-scalar task are depicted in the off-diagonal terms of Table~\ref{tab:results_seq2scal_cross}. First of all, these classification scores differ significantly from the diagonal and thus represent a degradation of prediction quality when using classifier $\mathfrak{C}_i$ for predicting squeal in data from system $j$ for $i \neq j$. Then, the scoring matrix is not symmetric such that classifier $\mathfrak{C_\mathrm{C}}$ performs with $\mathrm{MCC}=0.01$ on data set $B$, but $\mathfrak{C_\mathrm{B}}$ achieves a higher $\mathrm{MCC}=0.16$ on data set $C$. An interesting observation is that the pairwise similarity of the brake systems that have generated the data can be read from the results of the cross-evaluation. Intra-family scores, e.g. $A$ vs. $B$ and $C$ vs. $D$, are significantly lower than the diagonal entries, but still better than random prediction, i.e. $\mathrm{MCC}=0$. Inter-family scores, e.g. $A$ vs. $C$ and so forth, are essentially zero, indicating full randomness of the predictions. In a last step, all data are merged to a single set to fit a model on \textit{all} observations. Trained on $24755$ observations and validated against $10613$ observations, this model exhibits only poor scores that can be interpreted as some kind of average between the diagonal entries and the off-diagonal entries. The model performs still better than random, but is very unlikely to be used for squeal prediction in its current form.   
\begin{table}[h]
\centering
\caption{Results of the sequence-to-scalar classifiers: the main diagonal reports the validation results from Table~\ref{tab:results_seq2scal} in terms of the average MCC score. The off-diagonal entries report the MCC scores for the cross-evaluation study where a model trained on one data set is used to make predictions on another data set. Additionally, the results for a model trained and evaluated on all data at once is shown}
\begin{tabular}{c c c c c c c}
\toprule
& \multicolumn{5}{c}{model trained on} \\
 & &  $A$ & $B$ & $C$ & $D$ & all \\ 
\addlinespace
\parbox[t]{1mm}{\multirow{4}{*}{\rotatebox[origin=c]{90}{model evaluated on }}} & $A$ &  \cellcolor[RGB]{211,211,211} $0.78$ & 0.45 & -0.08 & -0.09 & - \\
\addlinespace
& $B$ & 0.46 & \cellcolor[RGB]{211,211,211}$0.72$ & 0.01 & 0.01 & - \\
\addlinespace
& $C$ & -0.02 & 0.16 &\cellcolor[RGB]{211,211,211} $0.65$ & 0.47 & - \\
\addlinespace
& $D$ & -0.2 & -0.01 & 0.43 & \cellcolor[RGB]{211,211,211}$0.64$ & - \\
\addlinespace
& all & - & - &- & - & $0.24$ \\
\addlinespace
\bottomrule
\addlinespace
\end{tabular}
\label{tab:results_seq2scal_cross}
\end{table}

For the sequence-to-sequence task, the results in Table~\ref{tab:results_seq2seq_cross} are qualitatively similar to the previous case. Off-diagonal scores for a data set from a different family of brake systems are very low and essentially indicate that the model prediction is no better than random. For similar brake systems, higher MCC values in the range between $0.37$ to $0.51$ are obtained. Again, we note that the actual value of the MCC cannot be compared to the scalar prediction case due to its formulation based on each time step. However, there is certainly room for improvement especially for the classifiers $\mathfrak{C}_{\mathrm{C}}$ and $\mathfrak{C}_{\mathrm{D}}$.   

\begin{table}[h]
\centering
\caption{Results of the sequence-to-sequence classifiers: the main diagonal reports the validation results from Table~\ref{tab:results_seq2seq} in terms of the average MCC score. The off-diagonal entries report the MCC scores for the cross-evaluation study where a model trained on one data set is used to make predictions on another data set. Additionally, the results for a model trained and evaluated on all data at once is shown}
\begin{tabular}{c c c c c c c}
\toprule
& \multicolumn{5}{c}{model trained on} \\
 & &  $A$ & $B$ & $C$ & $D$ & all \\ 
\addlinespace
\parbox[t]{1mm}{\multirow{4}{*}{\rotatebox[origin=c]{90}{model evaluated on }}} & $A$ &\cellcolor[RGB]{211,211,211} 0.78 & 0.46 & -0.04 &-0.04  & - \\
\addlinespace
& $B$ & 0.51 &\cellcolor[RGB]{211,211,211} 0.75 & 0.04  & -0.01 & - \\
\addlinespace
& $C$ & 0.03 & 0.15  &  \cellcolor[RGB]{211,211,211}0.62& 0.41 & - \\
\addlinespace
& $D$ & -0.08 & -0.01 & 0.37 & \cellcolor[RGB]{211,211,211}0.50 & - \\
\addlinespace
& all & - & - &- & - & 0.23 \\
\addlinespace
\bottomrule
\addlinespace
\end{tabular}

\label{tab:results_seq2seq_cross}
\end{table}

As a result, all  four brake systems studied in this work exhibit very individual instability regimes in the loading parameter space spanned by the available data. The neural networks employed here were not able to learn an underlying instability pattern that is invariant to the physical brake system realization. Considering the vast amount of research on brake squeal, its instability mechanisms and parameter sensitivities, this finding is not surprising, but rather expected. However, considering a single brake system, the high prediction quality obtained by recurrent networks is a promising finding. Locally, the instability conditions giving rise to squeal can be learned from rather small data sets in an engineering-compliant fashion, i.e. using data from conventional NVH testing. Larger data sets are likely to further improve the prediction quality of the network-based NVH twins developed in this work.

\section{Conclusion}

This work proposes to use data-driven analysis approaches to detect brake noises and to predict friction-induced noise occurrence of brake systems. A computer-vision inspired vibration detection and characterization method is discussed in the first part of this work. The comparison to hard-coded spectral algorithms reveals the large potential of deep learning approaches for vibration research. While the conventional approach exhibits superior performance in high-frequency squeal detection, the deep learning approaches allow to robustly detect other sounds and overall increase the detection quality for highly automated data processing. In data-intensive noise and vibration research activities, the present approach represents a flexible, fast and engineering-compliant solution. Using the data annotation methods form the first part, the second part illustrates how recurrent deep neural networks can identify loading patterns that are related to structural instability. For the first time, it has been shown that not only the emergence of self-excited squeal, but also the time instant of onset can be predicted by purely data-driven approaches. Using the chosen model configuration, it does not seem to be advisable to train a single model on multiple data stemming from different braking systems. Instead, individual classifiers for individual brake systems can be a promising starting point for further research on instability patterns encoded in the loading conditions monitored during NVH testing. 

This work presents a first proof-of-concept for using data-driven methods and deep learning in brake squeal research. Further studies are required to confirm the findings presented here using four relatively small data sets. Future work will focus on different model architectures and hybrid methods including finite element formulations. Besides the occurrence of squeal, also the squeal level and the squeal frequency may be interesting characteristics to predict. Furthermore, the limited explainability of deep learning classifiers poses challenges for engineering design decisions. For example, higher degrees of explainability may support the design of new friction materials to reduce squeal. Also, it would be of interest to see whether the same built of a brake system always generates the same instability regimes or whether the variability is system dependent, or relies on the small differences the brake system is assembled or mounted. Overall, our findings are promising and may foster new data-driven research on friction-induced dynamics of mechanical structures.

 \section*{Acknowledgment}
MS was supported by the German Research Foundation (DFG) within the Priority Program "`calm, smooth, smart"' under the reference Ho 3851/12-1.

\section*{Author contributions}

\textbf{Merten Stender:} Conceptualization, Methodology, Data Curation, Software, Formal analysis, Investigation, Writing - Original Draft, Visualization, Writing - Review \& Editing  \textbf{Merten Tiedemann:} Validation, Formal analysis, Investigation, Resources, Data Curation, Writing - Original Draft, Project administration \textbf{David Spieler:} Methodology, Validation, Writing - Review \& Editing, Supervision \textbf{Daniel Schoepflin:} Methodology, Software, Formal analysis, Data Curation, Writing - Original Draft, Visualization \textbf{Norbert Hoffmann:} Conceptualization, Investigation, Resources, Writing - Review \& Editing, Supervision, Project administration, Funding acquisition \textbf{Sebastian Oberst:} Conceptualization, Writing - Original Draft, Writing - Review \& Editing, Supervision

\newpage
\appendix

\section{Brake noise detection and characterization}
\subsection{Confidence scoring for spectral squeal detection}
\label{app:confidence_score_spectral}

\begin{figure}[htb]
\includegraphics[width=1.0\columnwidth]{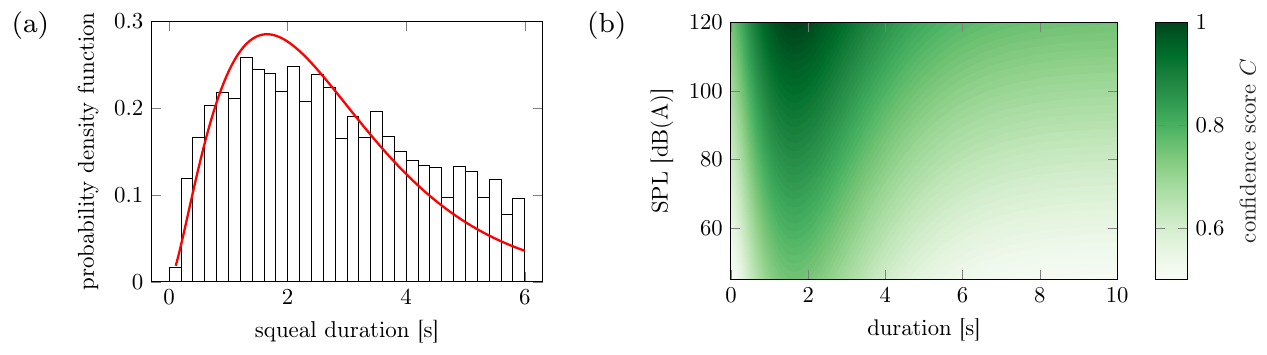}
\caption{Typical distribution of squeal duration (a) fitted by the Gamma probability density function \ref{eq:GammaPDF} and overall confidence score \ref{eq:confidenceScore} for the spectral detector as a function of squeal duration and squeal sound pressure level}
\label{fig:squealConfidence}
\end{figure}

To allow for a consistent comparison to deep learning noise detectors, we assign confidence values to the conventional spectral detector results. The squeal sound pressure level and the duration are used as proxies to estimate a confidence score. The higher the level, the more confident we are in the detection. Hence, the squeal level in the range of $\left[45, 120\right]$\,dB(A) is linearly mapped to the level rating metric $C_1 \in \left[0, 1\right]$. Using a histogram of approximately $4000$ noisy brakings, we study the distribution of squeal durations $d_{\mathrm{sq}}$ to assign a confidence score $C_2$ based on the duration of the detected event. It turns out that most of the squeals last for $1$ to $4$ seconds. Owing to its skew shape with single peak, the distribution is approximated by a Gamma probability density function (PDF)
\begin{equation}
y = f\left(x~|~a,b \right) = \frac{1}{b^a \Gamma \left(a \right)}x^{a-1} \mathrm{e}^{\frac{-x}{b}}
\label{eq:GammaPDF}
\end{equation}  
with shape parameter $a=2.55$ and scale parameter $b=1.07$, see Figure~\ref{fig:squealConfidence}~(a). The resulting PDF has a maximum value of $0.285$ which is used to scale the duration confidence such that $C_2 = f\left(d_{\mathrm{sq}}~|~a,b \right) \cdot \sfrac{1}{0.285} \in \left[0, 1\right]$. Now, a final confidence score $C$ is computed
\begin{equation}
C = \frac{C_1 + C_2}{4}+0.5 \quad \in \left[0.5, 1 \right]
\label{eq:confidenceScore}
\end{equation}
as a linear combination of the level and duration score. Figure~\ref{fig:squealConfidence}~(b) displays the confidence score value as a function of squeal duration and squeal level. This confidence score is a valid metric because we do not compare the confidence scores of the spectral and deep learning methods directly. The score is solely required for ranking the detections in the process of the PRC computation. Hence, the absolute values of this synthetic confidence score are technically irrelevant.

\subsection{Object detection metrics}
\label{app:prc}
\label{app:IoU}

To evaluate the correct bounding box size and location, the intersection over union (IoU) is computed. The IoU, see also Figure~\ref{fig:DLDetection}, measures the overlap of the ground truth $B_{\mathrm{gt}}$ and predicted $B_{\mathrm{p}}$ bounding box as the Jaccard index

\begin{equation}
\text{IoU} = \frac{\left | B_{\mathrm{gt}}\cap B_{\mathrm{p}} \right| }{\left | B_{\mathrm{gt}} \cup B_{\mathrm{p}} \right |} = \frac{\text{area of overlap}(B_{\mathrm{gt}}, B_{\mathrm{p}})}{\text{area of union} (B_{\mathrm{gt}}, B_{\mathrm{p}})} \quad \text{IoU} \in \left[ 0, 1 \right] \quad .
\label{eq:IoU}
\end{equation}

Taking into account a minimum IoU threshold and bounding boxes of a single class label, the validity of the object detection is defined as follows:
\begin{itemize}
	\item true positive (TP): $\mathrm{IoU} \geq \mathrm{IoU}_{\mathrm{threshold}}$
	\item false positive (FP): $\mathrm{IoU} < \mathrm{IoU}_{\mathrm{threshold}}$
	\item false negative (FN): ground truth objects for which there is no matching detection
\end{itemize}

There exist no notion of true negatives in object detection as this would correspond to labeling the background with an additional class \textit{background}. The $\mathrm{IoU}$ threshold depends on the specific metric definition and is usually set to $>50\%$. Now, each predicted bounding box $B_{\mathrm{p}}$ can be assigned to TP, FP or FN for each class individually. Typically, for each image the bounding boxes are sorted by their IoU score to cover special cases, such as multiple predictions for a single object. 

For object detection scoring, all predicted bounding boxes are listed in a table with their respective validity (TP, FP, FN). As each bounding box prediction features a confidence score from the detector model, the table can be ranked by this value. A minimal confidence score, e.g. $C_{\mathrm{min}}=0.8$, is set to only consider detections that have a higher confidence score than $C_{\mathrm{min}}$. Now, the cumulative precision and the recall values can be computed. The \textbf{precision-recall curve} (PRC) illustrates those cumulative measures to evaluate the performance of an object detector for each class separately. 

Precision, i.e. the positive prediction rate, and recall, i.e. the true positive rate, are contradicting metrics when considering different confidence levels. If the confidence threshold $C_{\mathrm{min}}$ is low, chances are high for over-prediction, i.e. high TP but also high FP rates. If a higher confidence level is considered, the number of false negatives will be high. A weak detector has to increase the number of detections to identify all relevant objects, thereby increasing the number of false detections, i.e. the false positives rate. Hence, a good object detector will maintain high precision and recall values for varying confidence scores, thereby finding a maximum of only relevant ground truth objects. This qualitative behavior can be measured by the area under the curve (AUC) of the PRC. The average precision (AP) \cite{Salton.1986} measures the mean precision for all recall values. In practice, the PRC is not monotonically decreasing, but has 'wiggles' caused by small ranking deviations of the samples, see Figure~\ref{fig:classification_PRC}. Therefore, the AUC is typically computed by interpolating the PRC. 11-point interpolation segments the recall into 10 equidistant intervals and samples the precision at the maximum precision value per interval, also called \textit{TREC sampling}. In this work we follow the PASCAL VOC definition of the AP \cite{Everingham.2015} which uses all points for an interpolation and estimation of the AUC. Averaging the APs over all classes gives the mean average precision mAP.  

For the deep learning models the minimum confidence level for reporting a predicted bounding box is set to $C_{\mathrm{min}}$. Lower confidence levels result in more false positive detections, while higher values lead to an increased number of false negatives. \ref{app:confidenceLevelStudy} demonstrates how and which confidence threshold values are selected for each class of brake noise.  

\subsection{Studies on the minimum confidence level}
\label{app:confidenceLevelStudy}

\begin{figure}[htb]
\includegraphics[width=0.99\columnwidth]{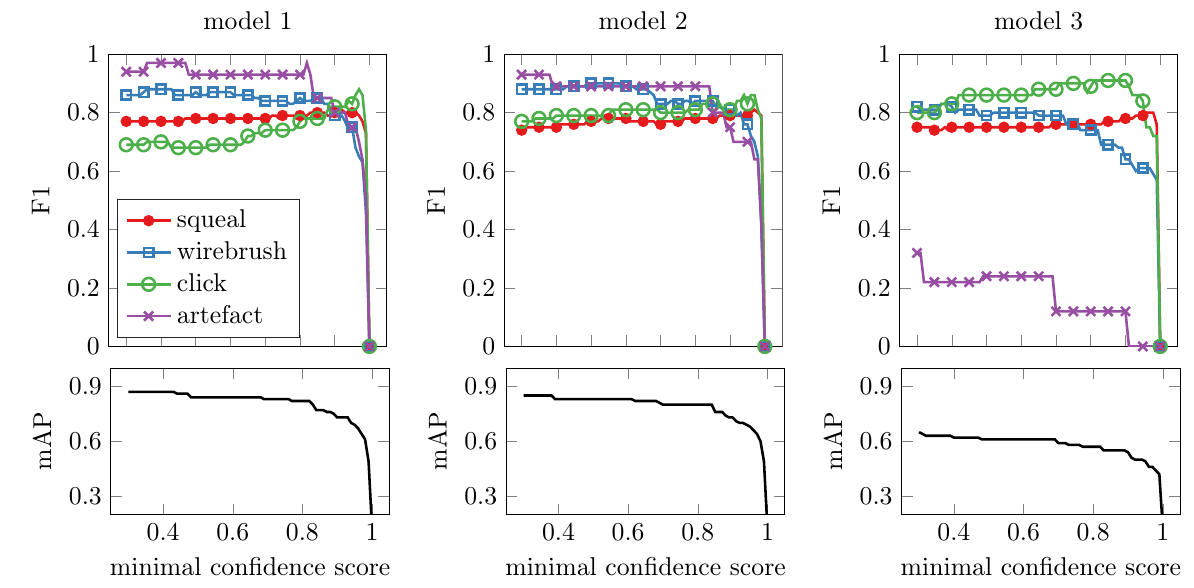}
\caption{Classification quality metric $F_1$ as a function of the minimal confidence score $C_{\mathrm{min}}$ required for reporting an object in the reference data set and resulting mean average precision score mAP}
\label{fig:classificationConfidence}
\end{figure}

The predictions of the deep learning object detector models come with a confidence score. For building an optimal brake NVH detection algorithm, we have to set a minimum confidence threshold $C_{\mathrm{min}}$ for reporting predicted objects as brake noise event. A small threshold will create too many false positives, while a too large threshold results in too many false negatives. Figure~\ref{fig:classificationConfidence} depicts the class-wise $F_1$ score for the classification task and the mean average precision score along increasing minimal confidence thresholds. Changing behavior can be observed for different models and individual object classes. To account for imbalanced representation of those classes in typical brake noise tests, we select the optimal confidence threshold such that high scores are achieved for squeal and a good balance is maintained for the remaining classes of other noises. Model 3 exhibits poor detection performance for the artefact class and weak performance for the wirebrush class. As a result, the mAP is significantly lower than for the other two models. These detectors show similar behavior for increasing confidence scores: the quality metrics rise for squeal and click sounds while the wirebrush and artefact classes show decreasing detection quality. In the range of $0.8 \leq C_{\mathrm{min}} \leq 0.9$ all $F_1$ scores are in the same order of magnitude. Hence, this parameter range is a valid choice for the minimal confidence score required for reporting predicted events when building an optimal brake NVH sound detector. The following thresholds $C_{\mathrm{min}}$ were selected: model 1 $C_{\mathrm{min}}=0.9$ (mAP$=0.73$), model 2 $C_{\mathrm{min}}=0.84$ (mAP$=0.8$) and model 3 $C_{\mathrm{min}}=0.88$ (mAP$=0.55$)    

\subsection{Object detection precision recall curves}
\label{app:objDetecPRC}

\begin{figure}[htb]
\includegraphics[width=0.9\columnwidth]{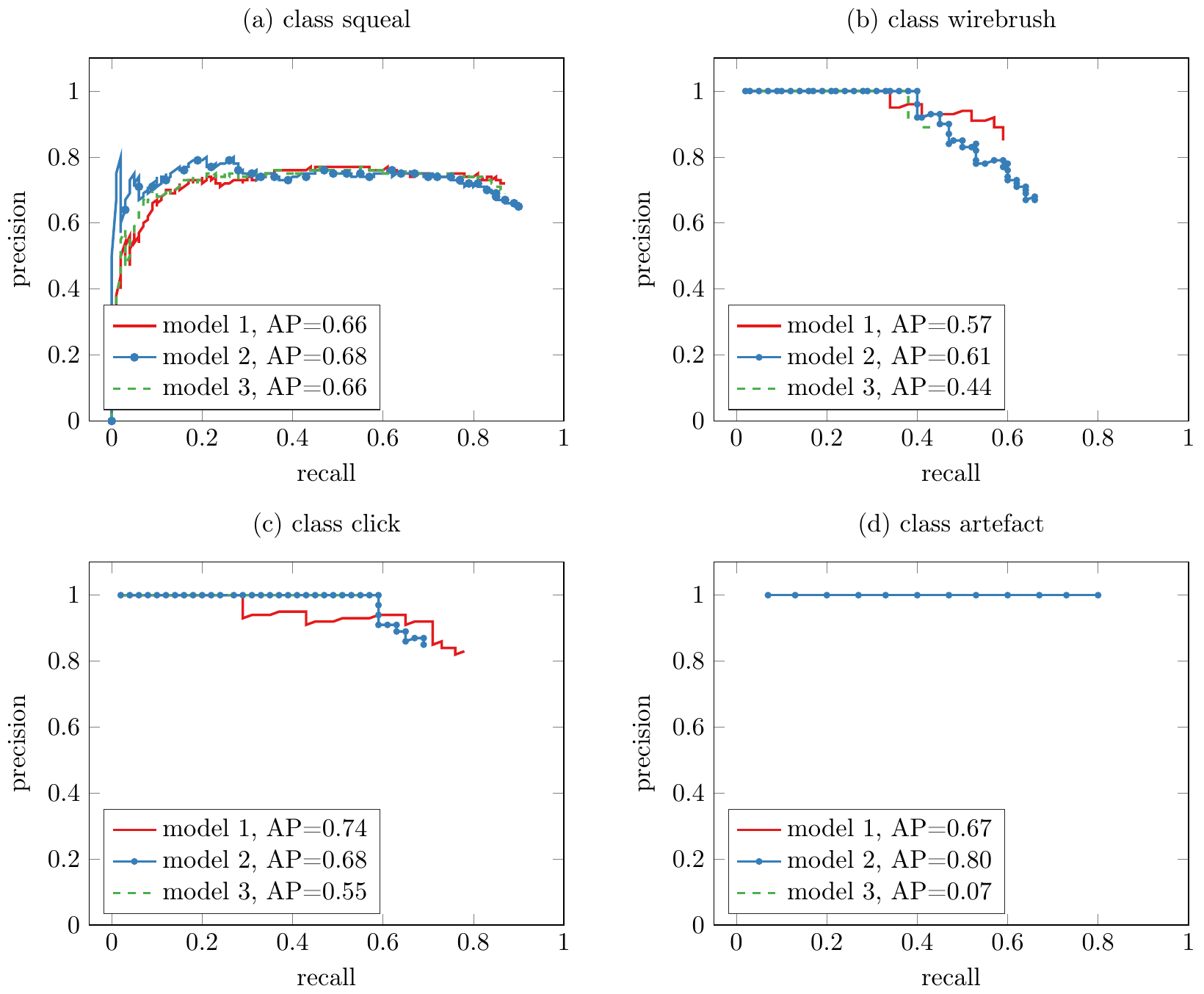}
\caption{Precision recall curves for the object detection task using a IoU threshold of $50\%$}
\label{fig:objectDetectionPRC}
\end{figure}

Figure~\ref{fig:objectDetectionPRC} depicts the PRC curves for the object detection task for each of the four brake noise classes and all three deep learning models. 

\newpage
\section{Brake squeal prediction}

\subsection{Data characteristics}
\label{app:data_char}

\begin{figure}[h]
\centering
\includegraphics[width=0.99\textwidth]{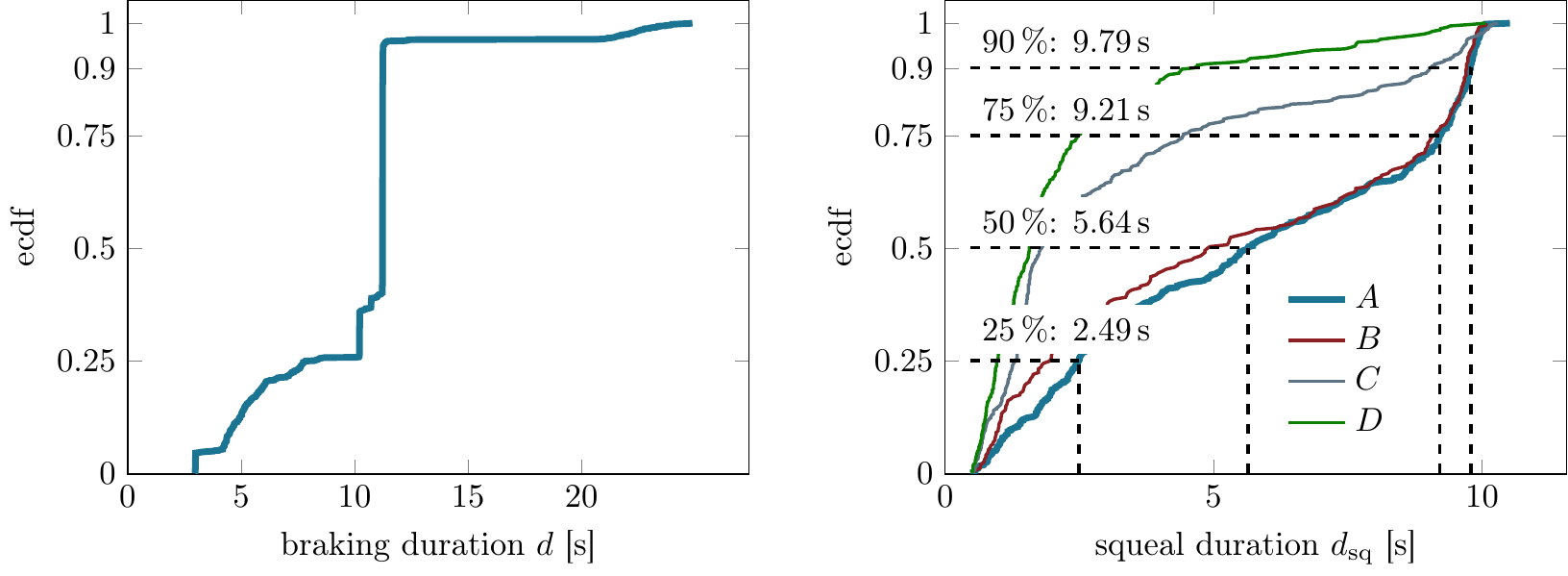}
	\caption[]{Left panel: empirical cumulative distribution function (ecdf) of the brake durations in data set $A$. The right panel depicts the ecdf of the squeal durations in data sets $A$, $B$, $C$ and $D$. Characteristic quantiles are marked for data set $A$ for which more than half of the squeals are longer than $5.5$\,s and $75\%$ of the squeals are longer than $2.49$\,s}
	\label{fig:datasetDurations}
\end{figure}

Figure~\ref{fig:datasetDurations} depicts the empirical cumulative distribution function of the braking durations and the squeal durations. While the braking durations are mostly prescribed by the testing procedure, the squeal durations arise from the self-excited instabilities and thus differ between the data sets. The two families of brake systems are clearly visible: systems $A$ and $B$ exhibit squeal sounds of similar duration distributions with more than half of the sounds being longer than $\approx 5.5$\,s. Systems $C$ and $D$ show significantly shorter squeal sounds. All relevant characteristics of the data sets are summarized in Table~\ref{tab:datasets}. The number of brakings and the squeal ratio are provided to give a general impression of the data sets and the uneven class distributions.  Figure~\ref{fig:instability} depicts those signals for a single braking in data set $A$. 

\begin{table}[h]
\centering
\caption{Characteristics of the data sets studied in this work: number of brakings $N$, number of squealing brakings $N_{\mathrm{sq}}$, squeal durations $d_{\mathrm{sq}}$ and their $25\%$ and $50\%$ quantiles as displayed in Figure~\ref{fig:datasetDurations}}
\begin{tabular}{c c c c c}
\toprule
data set & $N$ & $N_{\mathrm{squeal}}$ & $d_{\mathrm{sq}}(\text{ecdf}=0.25)$\,[s] & $d_{\mathrm{sq}}(\text{ecdf}=0.50)$\,[s] \\
\midrule
\addlinespace 
$A$ & 1206 & 487 & 2.49 & 5.64 \\
$B$ & 1206 & 227 & 1.97 & 4.93 \\
$C$ & 1206 & 347 & 1.32 & 1.80 \\
$D$ & 1889 & 237 & 0.97 & 1.57 \\
\bottomrule
\addlinespace
\end{tabular}
\label{tab:datasets}
\end{table}

\subsection{Deep learning model configuration}
\label{app:lstm_config}
The deep learning time series classifiers are built using the Python frameworks of TensorFlow and Keras. A coarse hyperparameter grid search is carried out for finding appropriate model architectures in terms of the number of hidden LSTM layers ($n_{\mathrm{layers}}\in \left[1,2\right]$), number of nodes per layer ($n_{\mathrm{units}} \in \left[ 64, 128, 256\right]$), optimizer (adam, stochastic gradient descend), and the batch size ($n_{\mathrm{batch}}\in\left[ 16, 64, 128, 256\right]$). Furthermore, the parameters of the sliding window ($w$, $h$) data preparation step are included in the grid search. To account for the rather small data sets, stratified three-fold cross-validation is employed to evaluate the generalization of the models on hold-out data. Stratified data splits ensure the same class distributions in the training and test set, i.e. the same under-representation of squealing brakings in the current studies. In $k$-fold cross validation, the original data are split into $k$ subsets. $k-1$ data sets are used to train a model, and the remaining data set is used as validation. This procedure is followed for all combinations of the $k$ data sets, and the resulting performance scores are averaged. In the hyperparameter study, each model is trained for $200$ epochs using binary cross-entropy as loss. Due to the high variability and nonlinearity of the brake squeal phenomenon, we expect rather complex networks to be necessary for successful prediction of the structural response. However, deep models exhibit the risk of overfitting. To prevent the latter, drop-out at a rate of $0.1$ is used for the LSTM cells. Fully connected (FC) layers with sigmoid activation are utilized to create the model output. The search is performed on data set $A$, and the selected architecture is re-used for all remaining data sets. Otherwise, comparison of performance for different data sets would not be possible in the cross-evaluation section. The final model parameters of the hyperparameter study for both classification models are given in Table~\ref{tab:model_specs}. Large batch size values of $256$ and $256$ LSTM units turn out to be the main driver for performance gain. Adding a second LSTM layer did not substantially improve the score, so a single LSTM layer is chosen. The sequence pre-processing parameters and the optimizer have only a secondary impact. Overfitting was not observed for any of the models in the hyperparameter study, see Figure~\ref{fig:traininghistory_seq2scal}.

\begin{table}[h]
\centering
\caption{Model configurations found in the hyperparameter study for the two classifiers. In both cases, a batch size of $n_{\mathrm{batch}}=256$ is chosen for the adam optimizer. The multivariate input sequences feature $m=8$ channels}
\begin{tabular}{c lccl} \toprule
model & layer & input shape & output shape & configuration\\
\midrule
\addlinespace
\multirow{6}{*}{\shortstack[l]{sequence \\to \\ scalar}} & LSTM & $\left[ m \times n_{\mathrm{t}} \times n_{\mathrm{units}} \right]$ & $\left[ m \times n_{\mathrm{units}} \right]$ & dropout$=0.1$\\
& FC & $\left[ m \times n_{\mathrm{units}} \right]$ & $\left[ n_{\mathrm{units}} \right]$  & ReLu activation\\
& FC & $\left[ n_{\mathrm{units}} \right]$ & $\left[ 1 \right]$ & sigmoid activation \\
\addlinespace
 & $n_{\mathrm{t}} =w = 200$ & & & \\
  & $h = 75\%$ & & & \\
  & $n_{\mathrm{units}} = 256$ & & & \\
\addlinespace
\midrule
\addlinespace
\multirow{6}{*}{\shortstack[l]{sequence \\to \\ sequence}} & LSTM & $\left[ m \times n_{\mathrm{t}} \times n_{\mathrm{units}} \right]$ & $\left[ m \times n_{\mathrm{t}} \right]$ & dropout$=0.1$\\
& FC & $\left[ m \times n_{\mathrm{t}} \right]$ & $\left[ n_{\mathrm{t}} \right]$  & sigmoid activation\\
\addlinespace
 & $n_{\mathrm{t}} =w = 400$ & & & \\
  & $h = 75\%$ & & & \\
  & $n_{\mathrm{units}} = 256$ & & & \\
\bottomrule
\addlinespace
\end{tabular}

\label{tab:model_specs}
\end{table}

\subsection{Sequence-to-scalar classifier training}
\label{app:lstm_training_history}
\begin{figure}[htb]
	\centering
\includegraphics[width=0.95\textwidth]{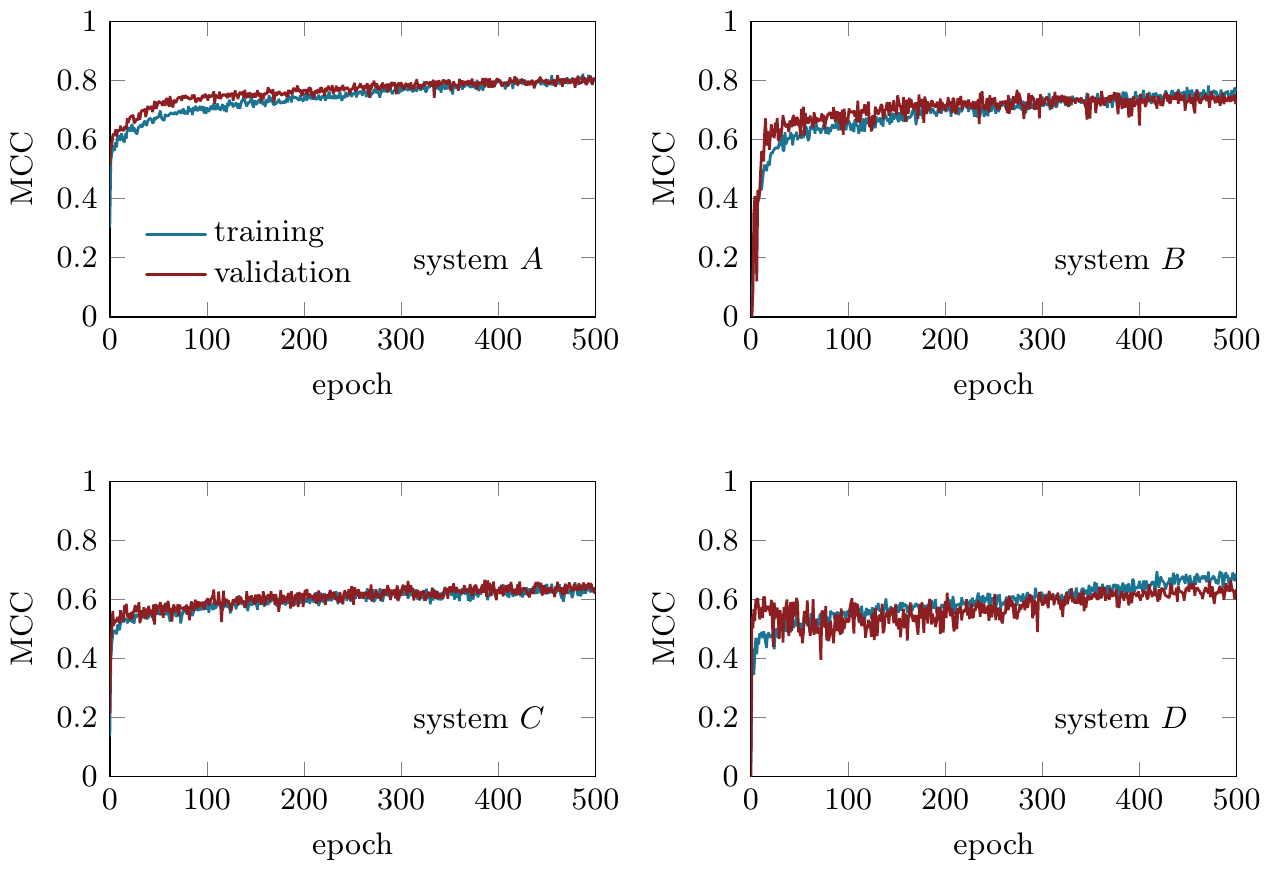}
	\caption[]{Training history of the sequence-to-scalar classifiers using the final model configuration and a stratified $70-30$ training-validation data split}
	\label{fig:traininghistory_seq2scal}
\end{figure}

Figure~\ref{fig:traininghistory_seq2scal} depicts the MCC classification scores along the training process of the sequence to scalar classifiers for the four data sets. As both the training and validation scores increase, the models are not overfitting. Further training could result in ever better performance values, but the risk of overfitting would increase, too. The classification score saturates for classifiers $A$, $B$ and $C$ while some performance gains can be expected for classifier $D$ when trained for more epochs 

\pagebreak

\bibliographystyle{elsarticle-num} 
\bibliography{deepLearning_brakeSqueal}




%
\end{document}